\documentclass{aa}
\usepackage[english]{babel}
\usepackage{amsmath}
\usepackage{xcolor}
\usepackage{graphicx}
\usepackage{txfonts}
\newcommand{\ind}[1]{_{\rm #1}}

\begin{document}

   \title{A charging model for the \emph{Rosetta} spacecraft}

   \author{F. L. Johansson
          \inst{1,2},
           A. I. Eriksson \inst{1},
           N. Gilet \inst{3},
           P. Henri \inst{3,4},\\
           G. Wattieaux \inst{5},
           M. G. G. T. Taylor \inst{6},
          C. Imhof \inst{7},
        \and
            F. Cipriani \inst{6}
          }

   \institute{Swedish Institute of Space Physics,
              Uppsala, Sweden\\
              \email{frejon@irfu.se}
         \and
             Uppsala University, Department of Astronomy and Space Physics, Uppsala, Sweden
            \and Laboratoire de Physique et Chimie de l'Environnement et de l'Espace, CNRS, Orl\'eans, France
            \and Laboratoire Lagrange, OCA, CNRS, UCA, Nice, France
            \and
            University Paul Sabatier Toulouse III, Toulouse, France
            \and
            ESA/ESTEC, Noordwijk, Netherlands
            \and
            Airbus Defence and Space GmbH, Friedrichshafen, Germany
             }

   \date{Received Jun 5, 2020; accepted Aug 6, 2020}
   
   \authorrunning{F. L. Johansson et al.}

\abstract
{The electrostatic potential of a spacecraft, $V\ind{S}$, is important for the capabilities of in situ plasma measurements. \emph{Rosetta} has been found to be negatively charged during most of the comet mission and even more so in denser plasmas.}
{Our goal is to investigate how the negative $V\ind{S}$ correlates with electron density and temperature and to understand the physics of the observed correlation.}
   {We applied full mission comparative statistics of $V\ind{S}$, electron temperature, and electron density to establish $V\ind{S}$ dependence on cold and warm plasma density and electron temperature. We also used Spacecraft-Plasma Interaction System (SPIS) simulations and an analytical vacuum model to investigate if positively biased elements covering a fraction of the solar array surface can explain the observed correlations.}
   {Here, the $V\ind{S}$ was found to depend more on electron density, particularly with regard to the cold part of the electrons, and less on electron temperature than was expected for the high flux of thermal (cometary) ionospheric electrons. This behaviour was reproduced by an analytical model which is consistent with numerical simulations.}
   {\emph{Rosetta} is negatively driven mainly by positively biased elements on the borders of the front side of the solar panels as these can efficiently collect cold plasma electrons. Biased elements distributed elsewhere on the front side of the panels are less efficient at collecting electrons apart from locally produced electrons (photoelectrons). To avoid significant charging, future spacecraft may minimise the area of exposed bias conductors or use a positive ground power system.}


   \maketitle

\section{Introduction}

The European Space Agency's (ESA) comet chaser, \emph{Rosetta,} monitored the plasma environment of comet 67P/Churyumov-Gerasimenko from August 2014 to September 2016. The scientific payload included the \emph{Rosetta Plasma Consortium} \citep[RPC,][]{carr_rpc:_2007}, dedicated to understanding the composition and evolution of the comet plasma. The RPC included, among other instruments, the \emph{Langmuir probe} \citep[LAP,][]{eriksson_rpc-lap:_2007,eriksson_cold_2017} and the \emph{Mutual Impedance Probe} \citep[MIP,][]{trotignon_rpc-mip:_2007,henri_diamagnetic_2017}. Because the instruments are mounted on \emph{Rosetta}, the RPC observations of charged particles are influenced by the electrostatic potential of the spacecraft with respect to its environment, $V\ind{S}$, but several RPC measurements can also be used to quantify this potential.

All objects in space exchange charge with their surroundings, mainly due to the collection of charged particles impacting the object and emission of electrons via the photoelectric effect and secondary emission. There are about as many negative electrons as positive ions in a plasma, but the electrons usually move much faster. In consequence, more electrons than ions tend to hit an uncharged spacecraft, giving it a negative charge unless the plasma is so tenuous that photoelectron emission dominates. An equilibrium is reached when the spacecraft becomes so negative that most plasma electrons are repelled. When the dominating compensating current is photoelectron emission, the spacecraft potential $V\ind{S}$ of a conductive spacecraft becomes \citep{odelstad_measurements_2017}
\begin{equation}\label{eq:Vs}
    V\ind{S} \approx -T\ind{e} \log \left( C\, n\ind{e} \sqrt{T\ind{e}}\  \right),
\end{equation}
where $n\ind{e}$ is the number density the electrons, which are assumed to be a Maxwellian population of characteristic temperature $T\ind{e}$, given in eV, and $C$ is a constant not depending on the plasma properties. The quantity in the logarithm essentially is the electron flux, which, together with $T_e$ is thus expected to drive the  $V_S$ in this case. If the collection of ions is a significant contribution to the current, the dependence on density becomes weaker.

\begin{figure*}
        \includegraphics[width=0.95\linewidth]{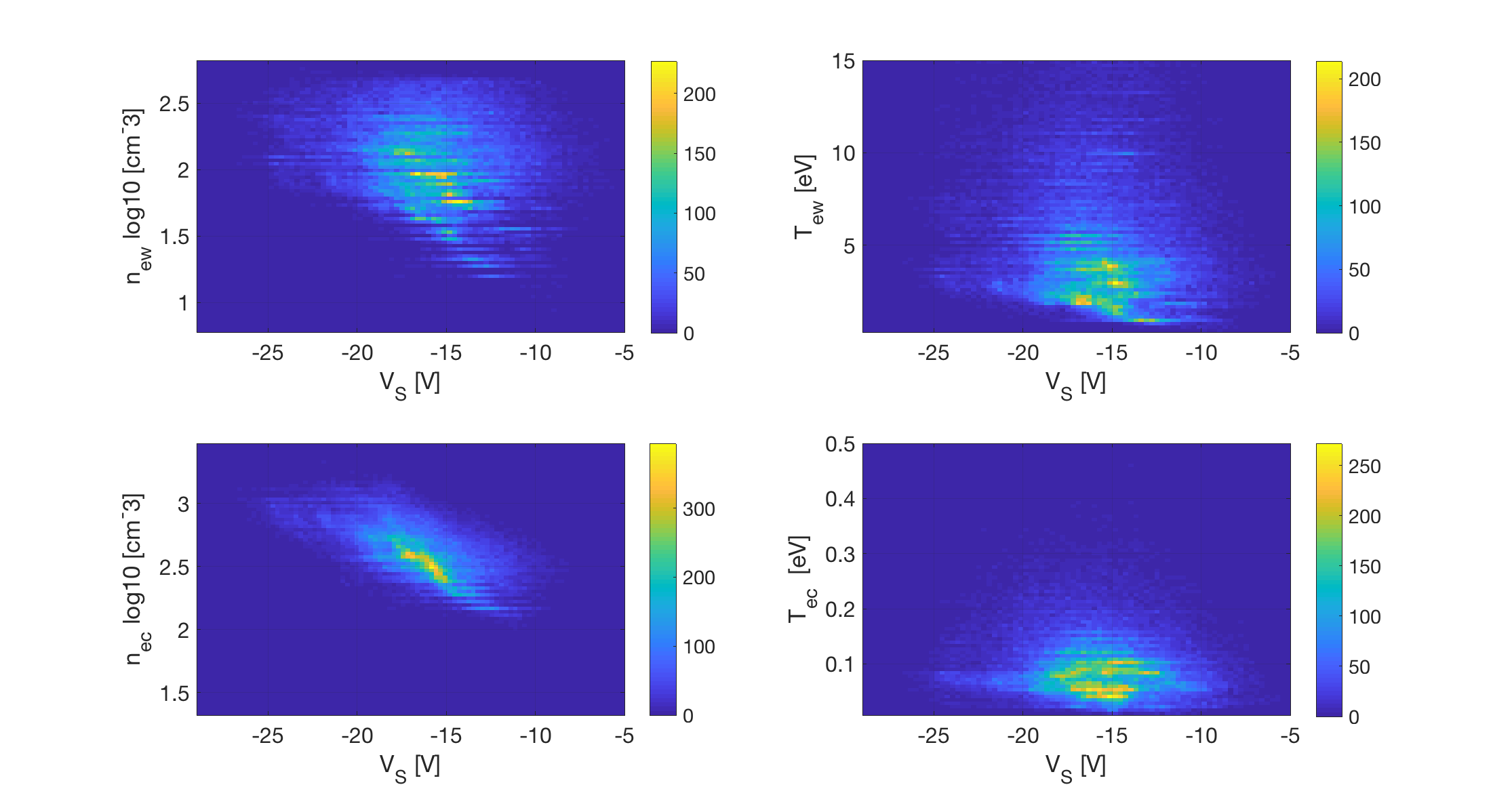}
    \caption{ 2D histogram in 80x100 bins of 88000 events of simultaneously measured spacecraft potential versus electron density (\textbf{left column}) and temperature (\textbf{right column}) from January to September 2016 at 2 to 3.8~AU. The identified electron populations by MIP from \citet{wattieaux_plasma_2020} are separated by temperature as warm ($T\ind{ew}$ $\approx 4$~eV,\textbf{top row}) and cold ($T\ind{ec}$ $\approx 0.1$~eV, \textbf{bottom row}). In total, 9700 outliers with either $T\ind{ew}$>15~eV  (9400 outliers) or  $T\ind{ec}$>0.5~eV (2800 outliers) have been removed.}
    \label{fig:vsc_and_gilet}
\end{figure*}

Predictions for the spacecraft potential of \emph{Rosetta} had already been produced  prior to launch. In two coupled studies, \citet{roussel_study_2004} used numerical simulations and \citet{berthelier_study_2004} investigated a spacecraft model in a laboratory plasma. Several plasma cases, including fully cooled (0.005~eV) cometary electrons were considered in the numerical simulations. Some simulations let the solar arrays to float into their own equilibrium potential which was found to be beneficial for reducing the otherwise often several volts positive potential observed in the simulations. The laboratory studies, therefore, emulated this case, with different surfaces on the model spacecraft insulated from each other and thus attaining their own equilibrium. The studies, which did not include biased elements on the solar arrays, suggested \emph{Rosetta} would attain potentials between a few times $-k\ind{B}T\ind{e}/e$ and about $+10$~V.

However, the spacecraft potential was continuously measured by LAP and found to be negative during most of the of \emph{Rosetta} cometary operations \citep{odelstad_evolution_2015,odelstad_measurements_2017}. The spacecraft often reached negative potentials around and in excess of -15~V, which have a severe effect on in situ measurements of the plasma environment surrounding the spacecraft as electrons are repelled \citep{eriksson_cold_2017} and positive ions are perturbed \citep{bergman_influence_2019,bergman_influence_2020}. From this spacecraft potential result, \citet{odelstad_measurements_2017}, with Eq.~(\ref{eq:Vs}), argued that the component dominating the electron flux is a thermal $\approx 5-10$~eV population omnipresent in the parts of the comet coma visited by \emph{Rosetta}. The existence of these warm electrons has also been verified by direct observation, by LAP \citep{eriksson_cold_2017}, by MIP \citep{wattieaux_plasma_2020}, as well as by the RPC Ion and Electron Sensor  \citep{broiles_statistical_2016}. However, there is also evidence of a highly variable cold ($\lesssim 0.1$~eV) population of electrons, independently detected by LAP \citep{eriksson_cold_2017,engelhardt_cold_2018} and MIP \citep{Gilet2019cold,wattieaux_plasma_2020}.

The cold electron population accounted for a significant, sometimes dominant, part of the total the electron density from January to September 2016 \citep{wattieaux_plasma_2020}, but due to its low temperature, the cold electron flux is low and is not expected to drive the spacecraft potential. However, in our analysis of LAP and MIP data during the cross-calibration activities for the final data deliveries for the ESA Planetary Science Archive, we came to notice a strong correlation between total plasma density, including the cold population and the spacecraft potential. Here, we report on these findings and present our investigation into why this is the case.

We structure this paper as follows: in Section~\ref{sec:data_analysis}, we present new statistics of simultaneously measured electron temperature, density, and spacecraft potential data showing unexpected correlations. To explain the results, we investigate details of the \emph{Rosetta} electrostatic design in Section~\ref{sec:hypothesis} and present a series of particle in cell simulations of a simplified model dealing with exposed biased elements on the spacecraft solar array in Section~\ref{sec:pillow} and discuss this model's shortcomings and merits. To improve our model, we adapt an analytical model of the vacuum potential of a charged disk in Section~\ref{sec:analytics} and run numerical simulations (Section~\ref{sec:spis_disks}) of a concentric disks geometry in an effort to highlight spacecraft design decisions with a critical influence on the cold electron current collection. Finally, we suggest a simplified model describing the \emph{Rosetta} current balance by setting up a system of equations to describe the current to the spacecraft and a positively biased surface behind a negative potential barrier in Section~\ref{sec:finalRosettamodel}, solve it numerically, and compare it to \emph{Rosetta} results.

\section{Data analysis} \label{sec:data_analysis}

A reworked analysis of MIP spectra with signatures of two electron populations \citep{wattieaux_rpcmip_model_2019,Gilet2019cold} yields an unprecedented precision in both energy and density of the thermal ($\approx 5$~eV) and cold ($\approx 0.1$~eV) electron populations. These estimates, combined with the recently published and improved spacecraft potential estimates from LAP (largely based on measurements published in \cite{odelstad_measurements_2017}) in AMDA (\url{http://amda.cdpp.eu/})
, give us simultaneous measurements of all parameters in Eq.~(\ref{eq:Vs}) for both populations.
We plot the MIP density estimates, as well as the mean of the LAP spacecraft potential estimates (typically 1 or 29 samples) taken during the acquisition period of each MIP spectra (typically 2~seconds) and plot them in Figure~\ref{fig:vsc_and_gilet}.

\begin{figure}
        \includegraphics[width=0.95\linewidth]{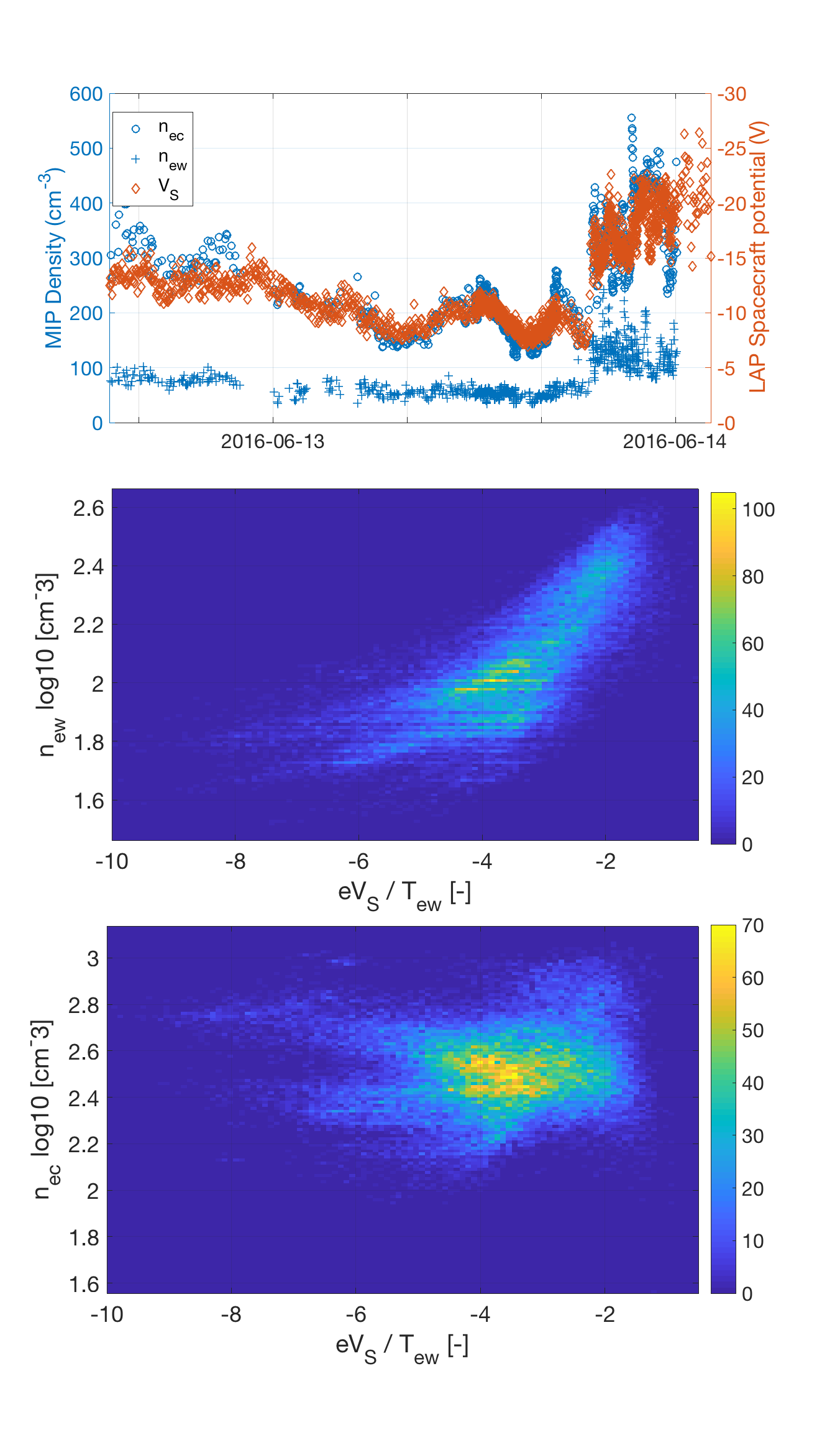}
    \caption{\textbf{Top:} Example time series of LAP spacecraft potential (diamonds),  cold (circles), and warm (pluses) electron density estimates from MIP in the same interval, exhibiting the strong correlation between spacecraft potential and cold electron density. \textbf{Middle:} Same data as shown in Figure~\ref{fig:vsc_and_gilet}, 2-D histogram of 120x150 bins of  $n\ind{ew}$ vs $eV\ind{S}/T\ind{ew}$. \textbf{Bottom:} As above, but with $n\ind{ec}$ on the y-axis.}
    \label{fig:vsc_and_gilet_times}
\end{figure}

In contrast to our model in Eq.~(\ref{eq:Vs}), the temperature variation in the two detected electron populations can only explain some of the variations in the \emph{Rosetta} spacecraft potential. Instead, the cold electron density has the clearest (logarithmic) relation to spacecraft potential out of the four parameters investigated.  However, for a uniformly charged spacecraft at -10~V, these cold (0.1~eV) electrons simply cannot reach the spacecraft and meaningfully contribute to the current balance that dictates the spacecraft potential.

The correlation between the cold electron density and the spacecraft potential is perhaps the clearest in a time-series, as plotted in Figure~\ref{fig:vsc_and_gilet_times} (Top), where we also observe a rather weak dependence on warm electron density to spacecraft potential. We also note that from 2016-06-12T16:00:00 to 2016-06-13T08:00:00, the average temperature of plasma electrons should increase (up to 50 percent) as the cold electron population density decreases, which according to our relation in Eq.~(\ref{eq:Vs}) would correspond to a more negative spacecraft potential. Instead, the opposite is true.

Normalising the spacecraft potential by $e/T\ind{ew}$ in Figure~\ref{fig:vsc_and_gilet_times} (middle panel) we see that an increase in warm electron density (for a fixed $T\ind{ew}$ does not drive the spacecraft more negative at all during the entire period from January to September 2016. Instead, it seems that an increase in $n\ind{ew}$ is associated with a decrease of $T\ind{ew}$, which is much more strongly coupled to the spacecraft potential. In general, it should not be surprising that in denser regions of the cometary ionosphere, the denser neutral cometary gas allows for more efficient cooling of all electrons \citep{edberg_spatial_2015}. What is also apparent is that the warm electron density does not strongly correlate with the cold electron density (bottom panel, same figure), which (albeit with more scatter) still shows the same trend of linearly increased charging with an exponential increase of cold electron density.

In the following section, we propose a mechanism to explain these observations.
\section{Exploring what drives \emph{Rosetta} to negative }\label{sec:hypothesis}

Ionospheric spacecraft have been observed to be driven negative by exposed, positively biased conductors on solar panels. For example, the OGO-6 satellite was observed to reach about -20~V \citep{Zuccaro1982a} and the International Space Station can reach as much as -140~V \citep{carruth_ISS_space_2012}. The reason is that such biased elements can draw a large electron current. To close the circuit, the spacecraft must respond by decreasing its potential to deflect more electrons away from it and to attract more ions from the plasma. This phenomenon can be regularly observed on small spacecraft equipped with Langmuir probes, where a large positive bias on the probe can result in a small negative shift of $V\ind{S}$ \citep[e.g.][]{ivchenko_disturbance_2001}. On \emph{Rosetta}, the surface area of about 80~cm$^2$ of each of the two LAP probes is negligibly small compared to the total spacecraft area (including solar panels) around 150~m$^2$ , so these cannot drive $V\ind{S}$ to the high negative values observed.%

\begin{figure}
        \includegraphics[width=1.0\linewidth]{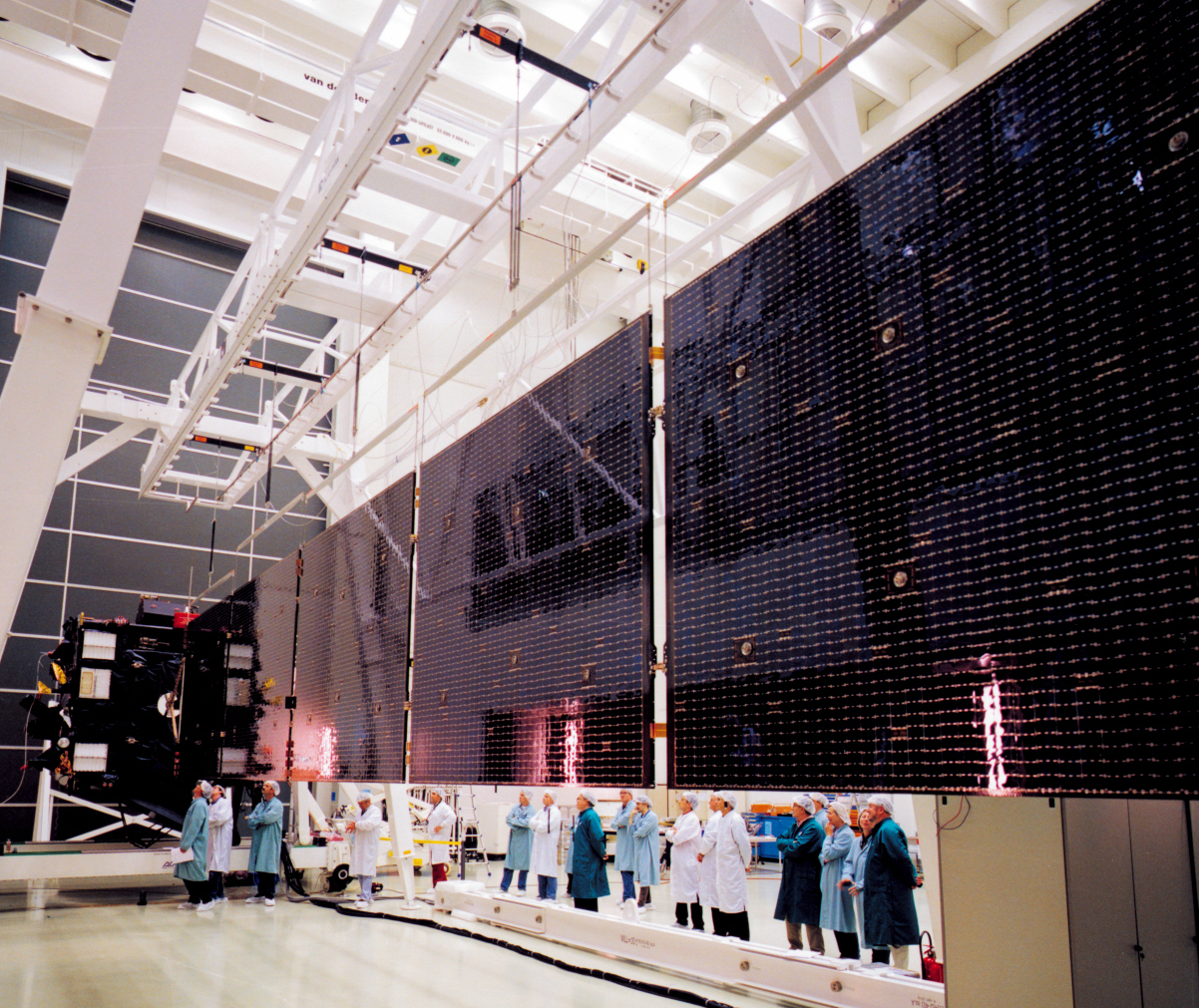}
    \caption{\emph{Rosetta} spacecraft and one of its solar wings. The solar cell cover glass on each cell is visible as small dark glossy surfaces, with metallic reflective interconnects above and below it. There are also 25 slightly larger reflective bus bar pairs, not to be confused with the six circular Kevlar cutter/hold-down points. Adapted from {\it Rosetta Solar panels} on ESA's website. Retrieved May 4, 2020, from \protect\url{sci.esa.int/s/w0e6nbW}. Copyright 2012 ESA–A, Van der Geest. Reprinted with permission.}
    \label{fig:rosetta}
\end{figure}

\begin{figure}
        \includegraphics[width=1.0\linewidth]{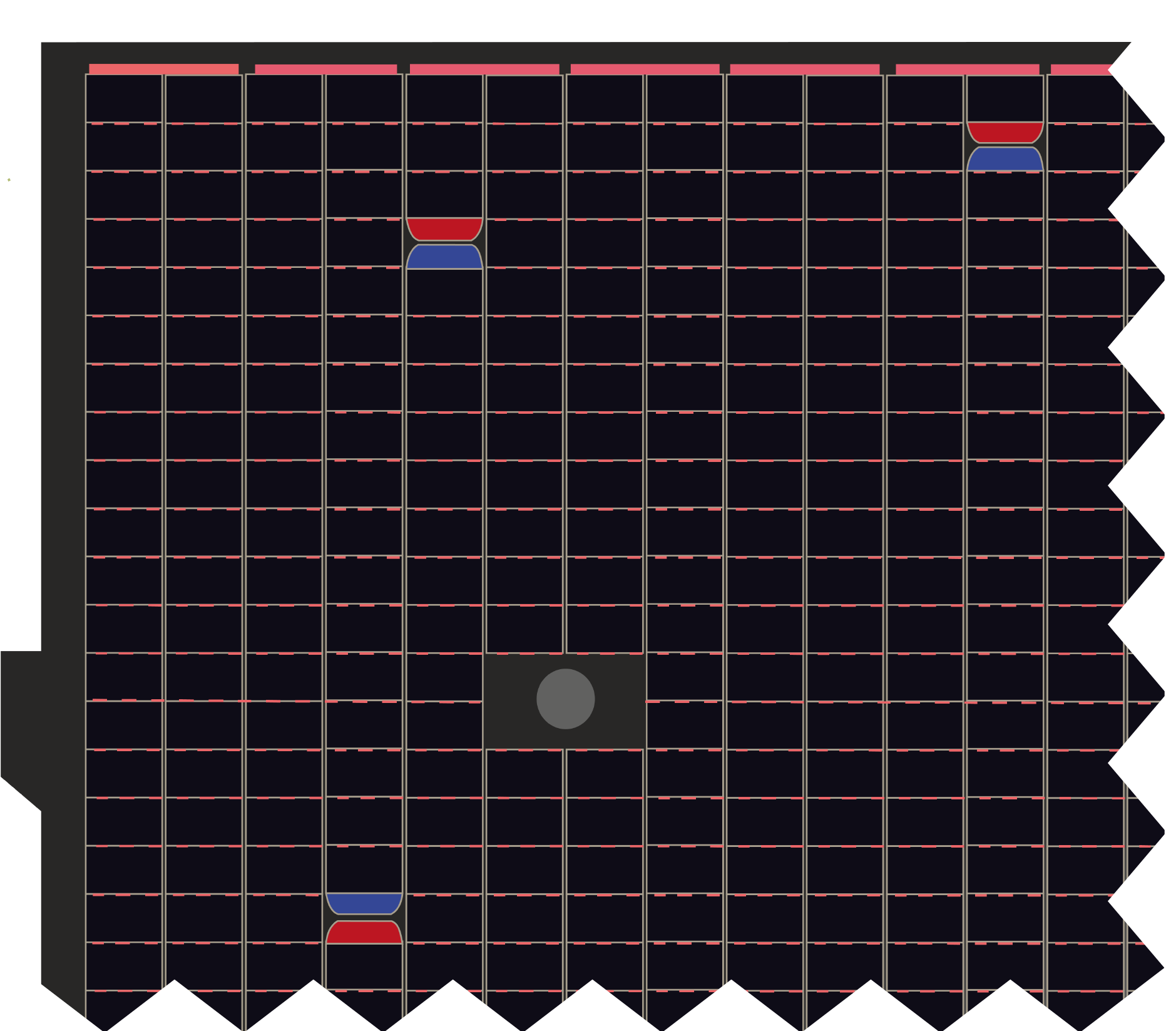}
    \caption{Artist impression of a corner section of the front side of a solar panel on \emph{Rosetta}. The black squares are individual solar cells covered with grounded cover glass, connected in series via (pink) interconnects in a column that wrap around to the next column near the top edge (and bottom, not shown) of the solar panel via a longer interconnect (also pink). At the start and end of each string of 91 solar cells are bus bars, marked with red (anode) and blue (cathode). The grey circle represents one of six circular Kevlar cutter/hold-down points. All surfaces coloured in pink and red are exposed positively biased conductors.}
    \label{fig:rosetta_panel}
\end{figure}

\emph{Rosetta} was not expected to be (and effectively never was) exposed to the large fluxes of high energy particles often driving spacecraft charging to dangerous levels in, for example, auroral zones \citep{eriksson_charging_2006,garrett_modeling_2008}. Efforts were  taken, nonetheless, to minimise exposed dielectrics and non-grounded conductors in order to provide a stable ground for plasma measurements. Providing a conductive and grounded outer layer was straightforward for most parts of the spacecraft. As \emph{Rosetta} was to be the first spacecraft operating on only solar power as far from the Sun as 5.25~AU, the front side of the large (64~m$^2$) solar arrays was a more complex issue, but in the end, it was decided that the solar cells would also be provided with cover glasses with a conductive indium tin oxide surface layer connected to spacecraft ground. In the low-energy dominated environments of concern here, the charging of dielectrics is not the primary concern. More of interest are exposed conductors, which can draw a significant current and, hence, influence the spacecraft potential. Thanks to the solar cell cover glasses and the equally conductive and grounded multi-layer insulation on the spacecraft body, the dominant fraction of all exposed conductive surfaces (we estimate at least 95\%) is at spacecraft potential. However, there are exceptions, particularly on the solar panels.

The \emph{Rosetta} solar array (Figure~\ref{fig:rosetta}) consists of 10 panels, each with 25 strings of 91 solar cells on its front side. While each cell has a conductive and grounded cover glass, there are small exposed biased conductors (interconnects) linking the cells in a string as well as the ends of a string to the spacecraft power bus. The single largest exposed positive potentials on a panel are the 25 small anodes of the bus bars at the end of each string, which can be seen as a sketch in Figure ~\ref{fig:rosetta_panel}. The bus bar anode is biased up to +79~V from spacecraft ground (and the bus bar cathode) on a string in open-circuit condition, and +65~V for a string operating at the maximum power point\footnote{The string voltage on the solar array is driven by many parameters, including the temperature and degradation of the cells and the power requested by the spacecraft}. The 89 interconnects in each string between the anode and cathode are therefore biased to an equidistant and linearly increasing potential for each consecutive solar cell in the string, such that the bias voltage on the last interconnect before a +79~V anode is +78.12~V, and the second to last is +77.24~V and so on. The bus bars are scattered on the panel, immediately surrounded by solar cells that are covered by a cover glass connected to spacecraft ground. The interconnects are more numerous but slightly smaller. Most of them are also scattered over the surface, but as each solar panel is organised into a grid of 57 rows and 42 columns, a string does not fit into one single column and so, it must wrap around when reaching a panel edge and continue along the next column. The upper and lower border of each solar panel front side are therefore lined by solar cell interconnects, and, as such, they are all biased to voltages between 0.7 and +78.12~V, depending on the bus bar anode potential.  A na\"{i}ve assumption might be to assume that incident electrons of any temperature could be collected at these voltages for the entire range of spacecraft potential measured in Figure
~\ref{fig:vsc_and_gilet} and, in some sense, eliminate the temperature dependence in Eq.~(\ref{eq:Vs}).

Based on simple Orbital-Motion Limited (OML) considerations, we see that small surfaces that are biased from the ground with a potential $V\ind{B}$ can easily dominate the positive current collection to a spacecraft as the current increases as a function of the absolute potential of the surface for any surface except an infinite plane \citep{laframboise_probe_1973}. For the simplest two-body problem of a spherical, positively charged body of surface area, $A,$ immersed in a plasma of density, $n$, the current collection of electrons is
\begin{equation}
    I\ind{e} = Ane\sqrt{\frac{k\ind{B}T\ind{e}}{2\pi m\ind{e}}}\left(1+\frac{eU}{k\ind{B}T\ind{e}}\right),
\end{equation}
where $U$ is the absolute (positive) potential of the body  $U = V\ind{B} + V\ind{S}$ and other constants have their usual meaning.
For 0.1~eV electrons and an exposed conductor at +75~V as discussed in the previous section, the current collection thus is leveraged by a factor of 750. Of course, charged elements on a spacecraft is a much more involved circuitry with a complex geometry that needs to be taken into account and requires numerical simulations.

\section{Numerical simulations} \label{sec:pillow}

The Spacecraft-Plasma Interaction System (SPIS) is a hybrid code package to simulate the spacecraft-plasma interaction, solving the Gauss's Law for electric fields, pushing particles, and simulating the spacecraft circuitry response and interaction with the plasma \citep{mateo-velez_spis_2012,mateo-velez_simulation_2015,sarrailh_spis_2015}. This work is a continuation of efforts of modelling the \emph{Rosetta} spacecraft in a cometary plasma environment by \citet{johansson_simulations_2016} and \citet{bergman_influence_2020} to understand the RPC instrument measurements. The simulation parameters for a reference simulation are provided in Table~\ref{tab:table}. The cometary ion population provides little current to the spacecraft system but ensures quasi-neutrality with a realistic \citep{stenberg_wieser_investigating_2017} but isotropic thermal velocity. To reduce the computational time of some of the SPIS simulations, we can simulate particles also as a Maxwell-Boltzmann fluid approximation instead of a full particle-in-cell (PIC) simulation. This treatment is not valid when there are positively biased elements present as the electron density in each simulation cell is extrapolated from the potential in that cell and, as such, we would overestimate the electron density near positive elements and within potential barriers. However, the reduction of computational (PIC) noise from a fluid approximation is very welcome for the purposes of demonstration.

\begin{figure}
        \includegraphics[width=1.0\linewidth]{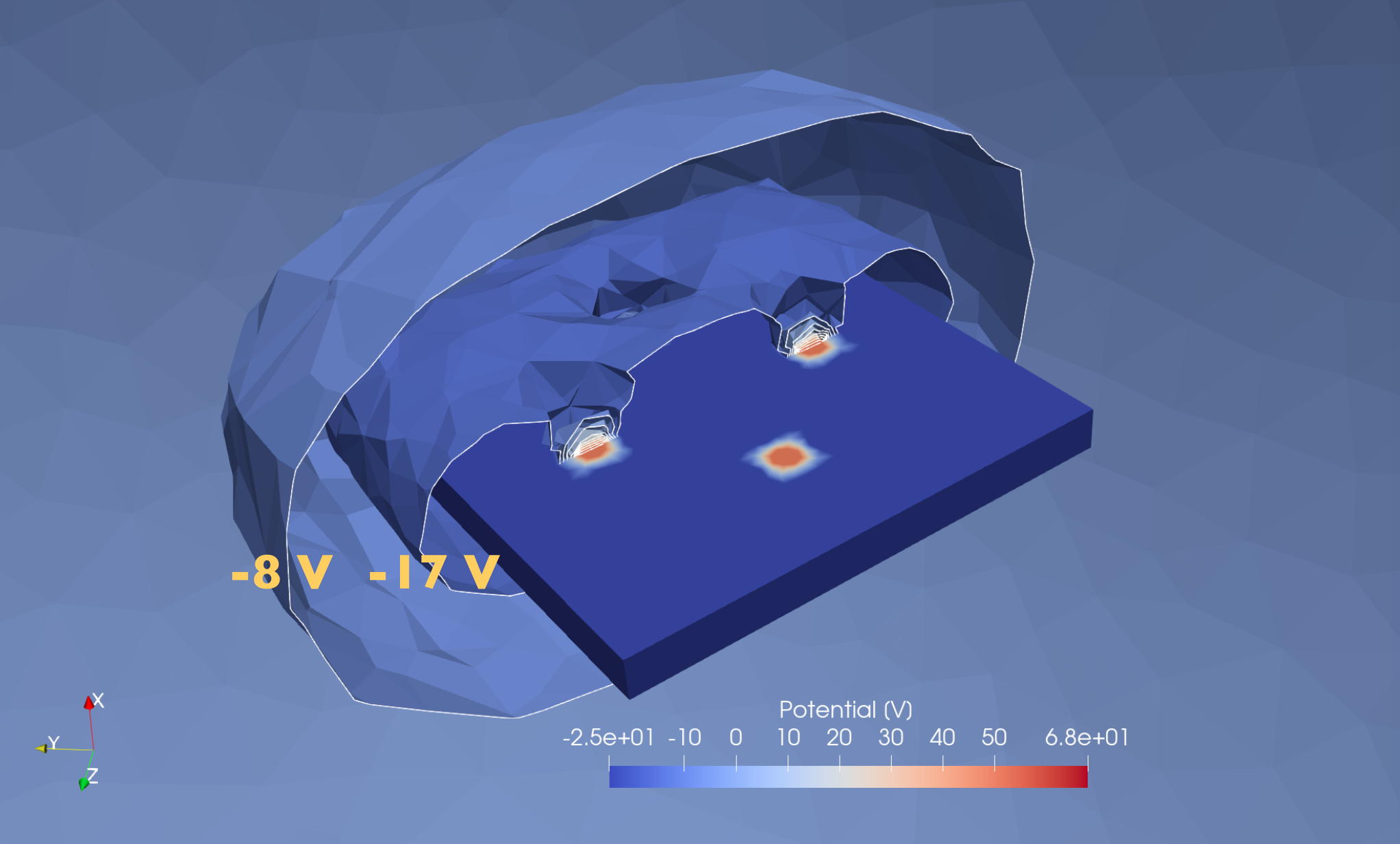}
    \caption{Visualisation of electrostatic potential struture from a SPIS simulation of a model with four $0.1\times0.1$~m +75V biased elements on a  $1.25\times1.25\times0.15$~m solar panel inside a spherical simulation volume of radius 15~m. Ten equipotential surfaces (cut in the Y=Z plane) from -17~V to +42.5~V are also plotted with the  -8~V and -17~V surfaces specifically labelled.}
    \label{fig:kudde}
\end{figure}

\begin{figure}
        \includegraphics[width=1.0\linewidth]{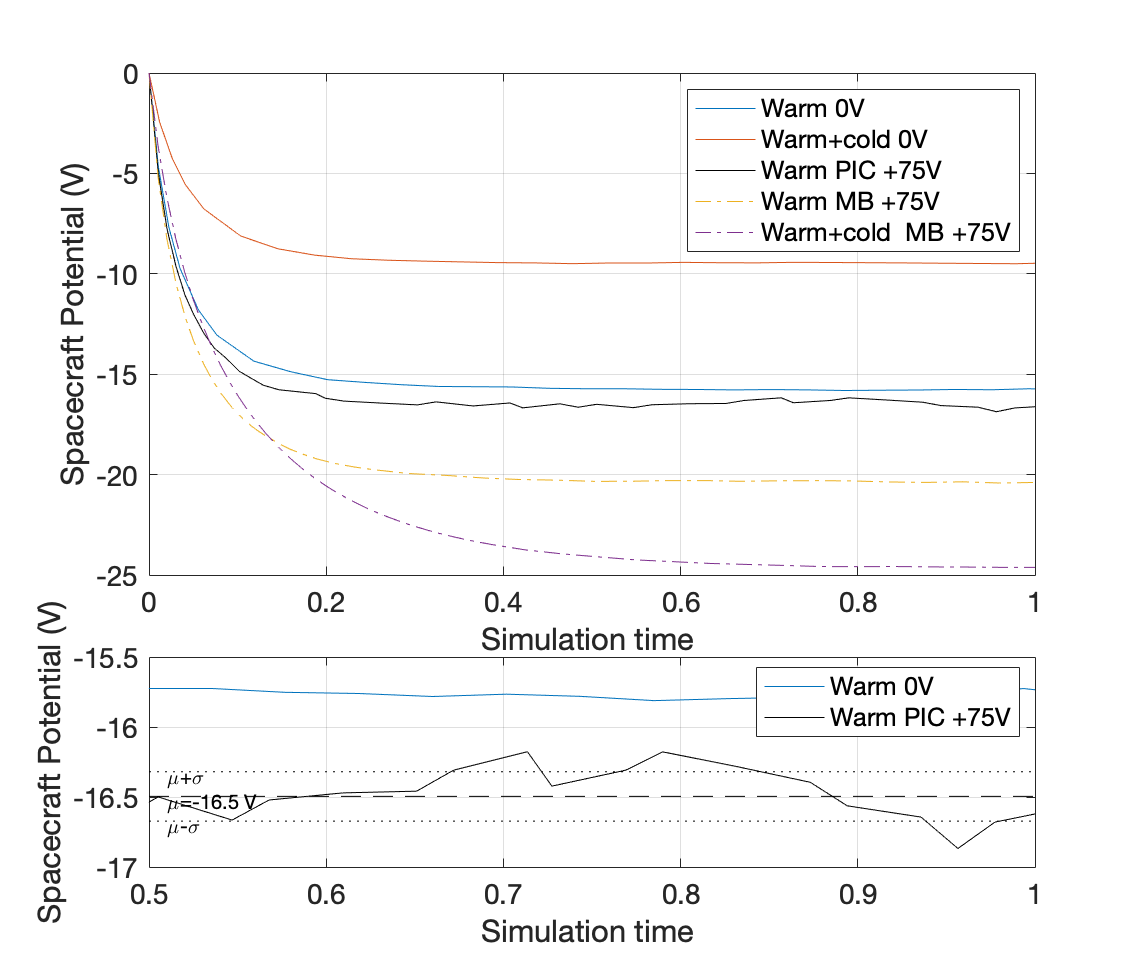}
    \caption{\textbf{Top:} Spacecraft (ground) potential evolution in five SPIS simulations. The reference simulation with no biased surfaces in a warm $T\ind{ew}=10$~eV plasma (blue line), with small charged surfaces of +75~V in a PIC simulation (black) or a fluid Maxwell-boltzmann simulation (yellow dot-dashed line). Also plotted, a simulation of the same plasma density but with 50 percent $T\ind{ec}=0.1$~eV electrons with either no biased surfaces (red), or with biased surfaces using the fluid approximation (purple dot-dash line). The Maxwell-Boltzmann simulations with surfaces at +75~V are strictly not valid but serve to illustrate the first approximation from Orbital-Motion-Limited theory. \textbf{Bottom:} Zoom-in of above, with the calculated mean (dashed black line) and a 1$\sigma$ range (dotted black lines) of spacecraft potential from the PIC simulation in this interval.}
    \label{fig:kudde_plot}
\end{figure}

We can also gain insights by studying simplified geometries; given the solar panels are the largest areas, we neglect the body and because each solar panel is large compared to the Debye length (which should be 30cm or less with cold electrons around), the whole array should essentially behave as a single solar panel. Therefore, we approximate \emph{Rosetta} with a box of size $1.25\times1.25\times0.15$~m, where the thickness is exaggerated for the ease of simulation but brings in only a negligible contribution to the current balance. We also include four $0.1\times0.1
~m$ symmetrically placed elements on the front side of the solar panel, which we set to a bias potential $V\ind{B} = 75$~V, as shown in Figure~\ref{fig:kudde}. All surfaces are simulated as indium tin oxide (ITO) for the purpose of photoemission and conductivity as it is the principal material on all sunlit surfaces on \emph{Rosetta} and we otherwise assume this to have a negligible effect on $V\ind{S}$ in a cometary (low-energy) plasma environment.

\begin{table}
      \caption{Table of reference SPIS simulation environment parameters used in Sections~\ref{sec:pillow} \& \ref{sec:spis_disks}.} \vspace{1em}
    \label{tab:table}
  \begin{center}
    \renewcommand{\arraystretch}{1.2}
    \begin{tabular}[h]{ll}
      \hline
          $n\ind{ew}$ & 100~cm$^{-3}$              \\
          $T\ind{ew}$            & 10~eV            \\
          $T\ind{i}$          & 1~eV    \\
          $m\ind{i}$          & 19~amu (H3O+) \\
          $u\ind{i}$          & 0~ms$^{-1}$    \\
         $j\ind{ph0}$ & 32 $\mu$Am$^{-2}$ \\
         $d_{\sun}$ & 3~AU \\
          Simulation Volume & Sphere of radius 15~m \\
          Secondary electron emission & Off\\
          Number of cells & $4.15\times 10^5$ \\
          Number of PIC particles & $\approx 5\times 10^6$ \\
      \hline \\
      \end{tabular}
  \end{center}
\end{table}

In Figure~\ref{fig:kudde}, we show an instructive example of the potential structure around a -25~V solar panel with small positively biased elements from a SPIS simulation with with the electron density set by a Boltzmann relation with the potential, complete with a three-dimensional potential barrier of -8~V.
As can be seen in Figure~\ref{fig:kudde_plot}, the effect of positively biased surfaces is two-fold:
\begin{itemize}
    \item The positively biased elements are collecting locally produced photoelectrons  from the surrounding surfaces as well, where the current magnitude as measured by SPIS corresponds to the photosaturation current of an area six times their size. Effectively, this turns photoemission off on an area six times as large as the positively biased elements on the solar array and drives the spacecraft potential to be more negative. For a more realistic case, with exposed biased elements that are spread over the entire (sunlit) panel, the photoemission of \emph{Rosetta} would be heavily suppressed. This can be part of the explanation on why \emph{Rosetta} was substantially negatively charged during the cometary mission and readily explains why \emph{Rosetta} only experienced moderate positive charging in the solar wind. \citep{odelstad_measurements_2017}.
    \item For standard OML, and indeed in the example SPIS Maxwell-Boltzmann fluid treatments in Figures~\ref{fig:kudde} and \ref{fig:kudde_plot}, the current to any positively charged surface (for\ the barrier potential, $U\ind{M}$) is severely exaggerated as most electrons born at a potential of 0~V at infinity cannot penetrate the barrier if their energy does not exceed $eU\ind{M}$. The aforementioned cold cometary electrons would contribute little to the current to these biased elements, as has indeed been confirmed by SPIS simulations with a PIC treatment of electrons.
\end{itemize}
This potential barrier effect is very effective in quenching the cold electron current when small positive elements are surrounded on all sides by grounded (negative) elements. Although the cold electron density population exhibits the exact behaviour we sought for in Section~\ref{sec:data_analysis} in the fluid approximation simulations, we must look for another explanation when a realistic treatment of electrons is applied.

As described in Section~\ref{sec:hypothesis} and in Figure~\ref{fig:rosetta_panel}, the interconnects are dispersed all over the solar panel surfaces, but they are (possibly crucially) always present at the top and bottom border of the solar panel front side, as the solar cell string wraps around to the next column. As for all interconnects on the solar array, on average, this border is expected to be biased between +30-40~V (although an average may not be the best descriptor for the net effect on current collection since many interconnects would be repelling electrons exponentially at $V\ind{B}<-V\ind{S}$) and can have less restricted access to electrons as it is not surrounded by negatively charged surfaces, an effect we investigate further in the following sections.

\section{Analytical model of solar panels with biased elements}\label{sec:analytics}

\subsection{Vacuum model for thin circular disk}\label{sec:sherman}

\begin{figure}
    \centering
    \includegraphics[width=0.8\columnwidth]{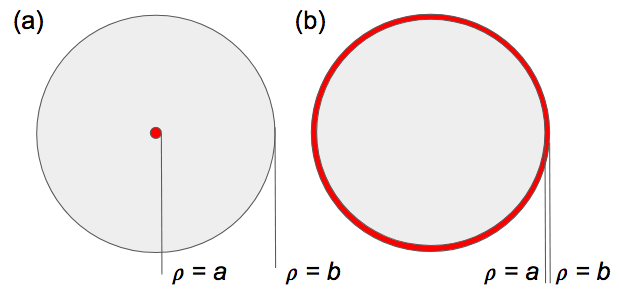}
    \caption{Geometry of the concentric disk model for modelling of solar panels with biased elements as described in Section~\ref{sec:sherman} for two different applications: Panel~(a) represents a single exposed biased conductor on the main area of the solar panel (Section~\ref{sec:interior}); panel~(b) exposes biased conductors along the edge (Section~\ref{sec:edge}). Grey areas represent the main solar panel at spacecraft potential, red a biased element.}
    \label{fig:disc-sketch}
\end{figure}

For a comparison and interpretation of the simulation results on the formation of potential barriers, we use an analytical solution of the Laplace equation around a thin circular disk which consists of two concentric parts, as illustrated by two examples in Figure~\ref{fig:disc-sketch}, an inner disc of radius, $a,$ at potential, $V\ind{in}$ , surrounded by an annulus of inner radius, $a,$ and outer radius, $b>a,$ at the potential $V\ind{out}$. At cylinder coordinates $(\rho, \phi, z),$ where $z=0$ defines the disc plane, \citet{Sherman1971a} found that the vacuum potential from this object is
\begin{align}\label{eq:sherman}
\Phi(\rho,z) & = V\ind{in} \frac{2}{\pi}
  \, \mathrm{atan} \left( \frac{b\sqrt{2}}{\sqrt{r^2-b^2+\sqrt{(r^2-b^2)^2 + 4z^2b^2}}}  \right) +\nonumber\\ & + (V\ind{out}-V\ind{in}) \frac{\sqrt{2}}{\pi}\ \cdot \nonumber\\ & \int_a^b  \sqrt{ \frac{r^2-s^2 + \sqrt{(r^2-s^2)^2+4z^2s^2} }{(r^2-s^2)^2+4z^2s^2} }  \cdot \frac{s}{\sqrt{s^2-a^2}}\ ds,
\end{align}
where $r^2 = \rho^2+z^2$. The integral can be analytically evaluated on the $z$~axis and in the disk plane $z=0$ to find that
\begin{align}
\Phi(0,z) & = \frac{2}{\pi} \left[ V\ind{in} \, \mathrm{atan} \frac{b}{z} + (V\ind{out}-V\ind{in}) \frac{z}{\sqrt{z^2+a^2}} \, \mathrm{atan} \sqrt{\frac{b^2-a^2}{z^2+a^2}}\ \right] \label{eq:Vz}
,\\
\Phi(\rho,0) & = \frac{2}{\pi} \left[ V\ind{in} \, \mathrm{atan} \frac{b}{\sqrt{\rho^2-b^2}} + (V\ind{out}-V\ind{in}) \, \mathrm{atan} \sqrt{\frac{b^2-a^2}{\rho^2-b^2}}\ \right]. \label{eq:Vrho}
\end{align}
We use these expressions to model potential barriers around solar panels with exposed biased conductors in Sections~\ref{sec:interior} and \ref{sec:edge} below. In extending an argument used by \citet{Sherman1971a} for the $z$~axis, the potential will have an extremum (minimum or maximum) on exactly one of the two axes. This is because at large distance the two plates look like a point with charge equal to their net charge, which must be either positive or negative. Far away, the potential decays as $\pm 1/r$, so if negative at large distance, the potential must have a minimum somewhere along the axis from the positively charged part, which is the $z$~axis if the positive part is the inner disk $\rho<a$ and otherwise the $\rho$~axis. Such a minimum in the potential is a maximum in electron potential energy and, hence, a potential barrier. By considering the net charge of the disks, the limit for barrier formation is found to be \citep{Sherman1971a}:
\begin{align}\label{eq:barrier_limit}
    \left| \frac{V\ind{out}}{V\ind{in}} \right| = \frac{1}{\sqrt{1-\left(\frac{a}{b} \right)^2}} - 1.
\end{align}
If the positive voltage, whether $V\ind{in}$ as in Figure~\ref{fig:disc-sketch}(a) or $V\ind{out}$ as in Figure~\ref{fig:disc-sketch}(b), is higher than allowed by this expression, there will be no barrier for electron collection by the positive element.

\subsection{Biased element in the centre of a solar panel}\label{sec:interior}

\begin{figure}
    \centering
    \includegraphics[width=\columnwidth]{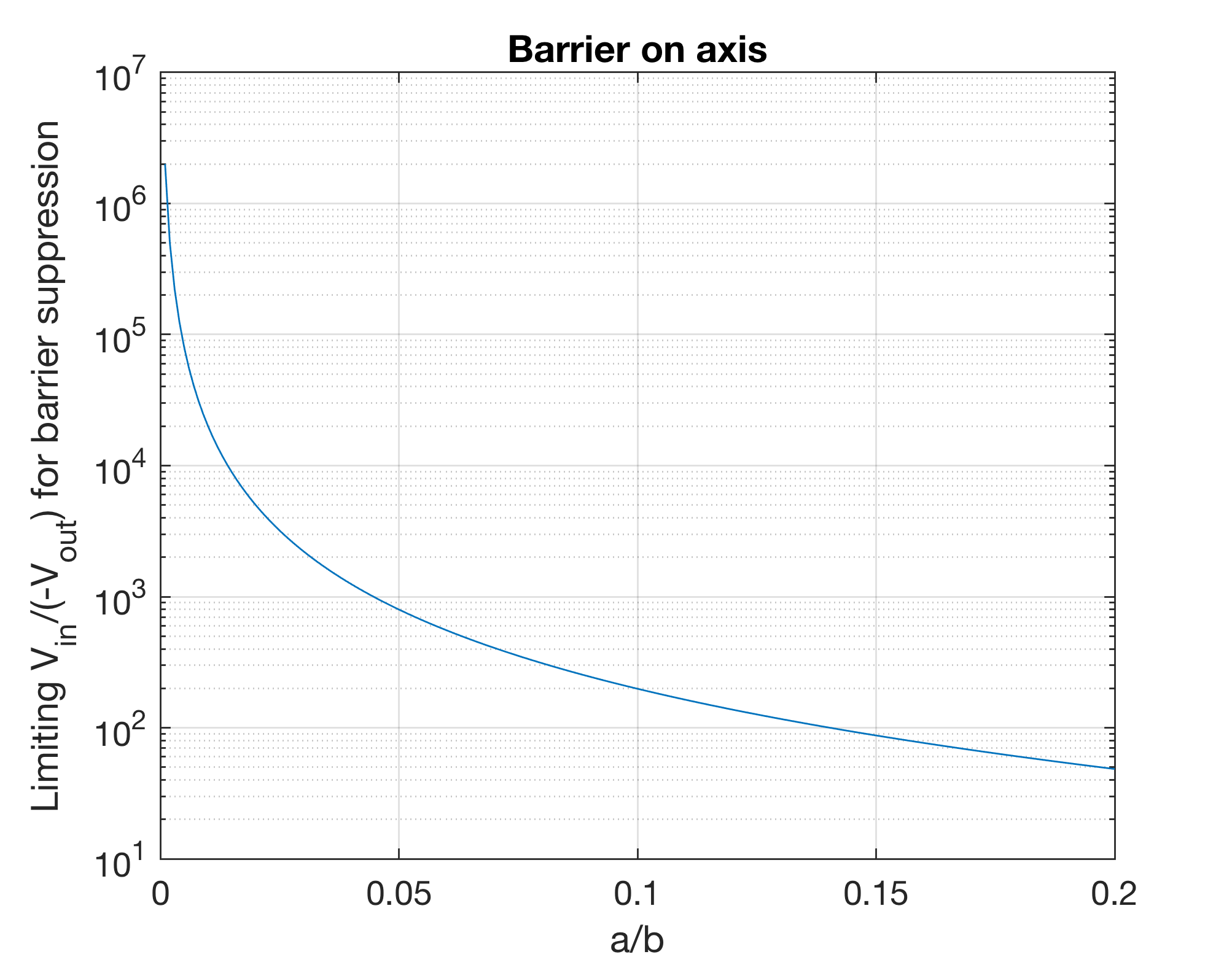}
    \caption{Limiting potential ratio for barrier suppression for a small biased element on the $z$~axis as given by Eq.~(\ref{eq:z_barrier_limit}).}
    \label{fig:z_barrier_limit}
\end{figure}

\begin{figure*}
    \centering
    \includegraphics[width=2\columnwidth]{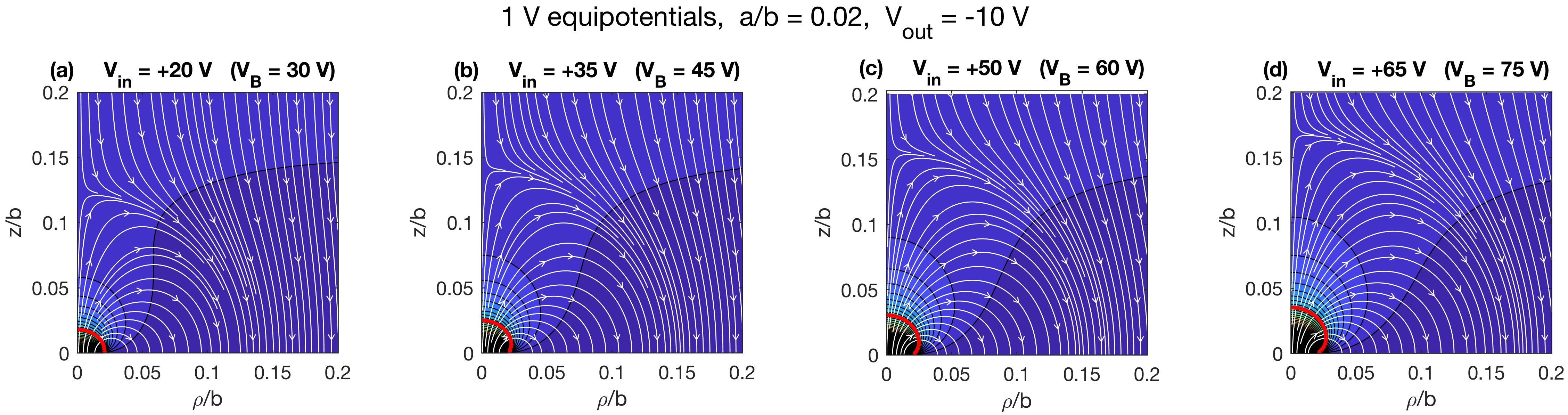}
    \caption{Vacuum potential and electric field pattern in the $\rho-z$~plane near the centre of a thin circular disk of radius $b$ as sketched in Figure~\ref{fig:disc-sketch}(a). The disk potential outside $\rho=a$ is $V\ind{in} = -10$~V while the potential $V\ind{out}$ in the centre $\rho < a$ varies as stated above each panel. In all cases, $a = 0.02\, b$. Numerically integrated electric field lines are plotted in white. The 0~V equipotential is shown in thick red; the potential is zero also at infinity. Black curves indicate equipotentials at every integer value (in volts), with the background colour further highlighting the potential.}
    \label{fig:sherman_contourf_button}
\end{figure*}

In this case, the outer annulus represents the solar panel at potential $V\ind{out} = V_S < 0$ and the inner disk represents the biased element at $V\ind{in} = V_S + V\ind{B}$. The minimum value of $V\ind{in}$ to break the barrier follows from Eq.~(\ref{eq:barrier_limit}) as
\begin{align}\label{eq:z_barrier_limit}
    \frac{V\ind{in}}{-V\ind{out}} = \frac{1}{\frac{1}{\sqrt{1-\left(\frac{a}{b} \right)^2}} - 1}.
\end{align}
Figure~\ref{fig:z_barrier_limit} shows this expression evaluated for a range of the radial ratio $a/b$. It is clear from this plot that forbiddingly large values of the bias ratio are needed for breaking the barrier, reinforcing the conclusion in Section~\ref{sec:pillow} that small positive elements on the interior of a solar panel would not collect cold plasma electrons.

In Figure~\ref{fig:sherman_contourf_button}, we show the vacuum electrostatic field near the centre of the same disk as calculated from the full expression Eq.~(\ref{eq:sherman}) for four different bias voltages $V\ind{B} = V\ind{in}-V\ind{out}$. The zero volt equipotential (red) ends at the intersection $\rho = a = 0.02\, b$ between the disks. All field lines starting on the positively biased inner surface ends up on the main solar panel area, as is expected since there is a potential barrier. The radius $\rho_0$ delimiting field lines connecting to the inner disk or to infinity can be seen to expand from about $0.12\, b$ to $0.18\, b$ as $V\ind{B}$ increases from $30$~V to $75$~V. If photoelectrons were massless and emitted from the solar panel in the normal direction with zero speed, they would follow the electric field lines. Using $\rho_0$ as a measure of the region from which photoelectrons from the solar panel would be collected by the biased element at the centre, we find that the area of this region is about 35 times bigger than the biased element itself already for $V\ind{B}=30$~V. While parameters are not perfectly comparable, this is still significantly more than the factor of about six that we found in the simulations in Section~\ref{sec:pillow}. This is expected, as photoelectrons would follow field lines perfectly only if massless and emitted at zero speed, neither of which is the case. Furthermore, our analytic model only considers a vacuum.

\subsection{Barrier potential around solar panel edges}\label{sec:edge}
\begin{figure*}
    \centering
    \includegraphics[width=2\columnwidth]{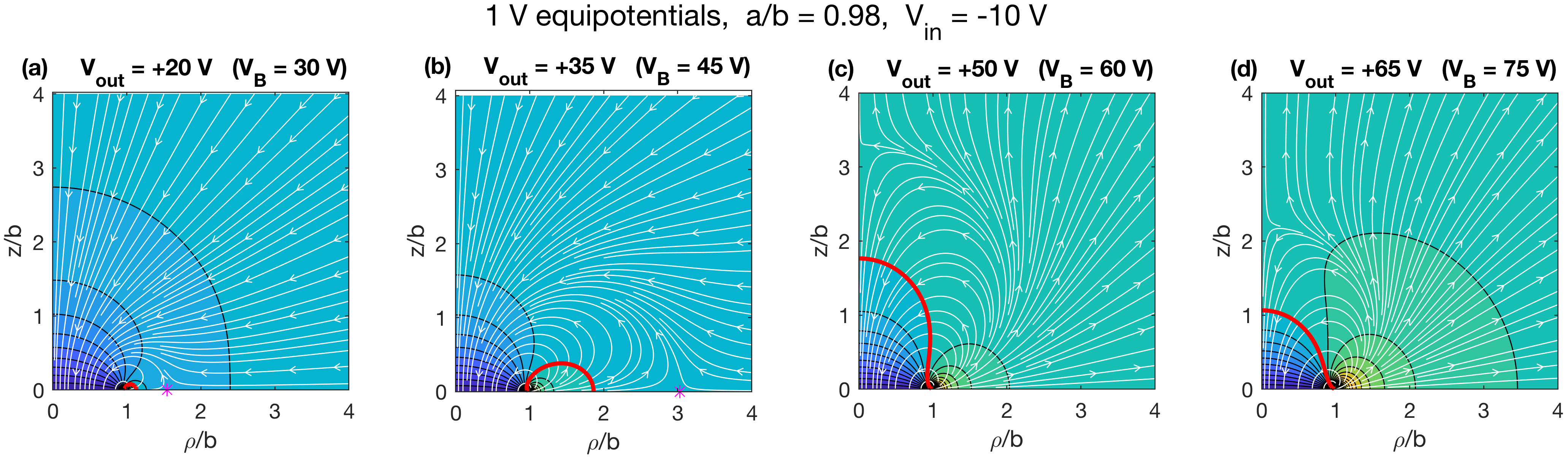}
    \caption{Vacuum potential and electric field pattern in the $\rho-z$~plane around a thin circular disk of radius $b$ as sketched in Figure~\ref{fig:disc-sketch}(b). The disk potential inside $\rho=a$ is $V\ind{in} = -10$~V while the potential $V\ind{out}$ in the annulus $a < \rho \leq b$ varies as stated above each panel. In all cases, $a = 0.98\, b$. Numerically integrated electric field lines are plotted in white. The 0~V equipotential is shown in thick red; the potential is zero also at infinity. Black curves indicate equipotentials at every integer value (in volts), with the background colour further highlighting the potential. The magenta star indicates the location of minimum electron barrier height.}
    \label{fig:sherman_contours}
\end{figure*}
We now turn our attention to the positively biased elements around the solar panel edges and apply our analytical vacuum model to this case. If the barrier effect is as effective here as we found for biased elements on the main solar panel surface, we could conclude that the biased elements on the solar panels cannot be responsible for the strongly negative potential of \emph{Rosetta}. Here we consider whether this is indeed the case.

\begin{figure}
    \centering
    \includegraphics[width=\columnwidth]{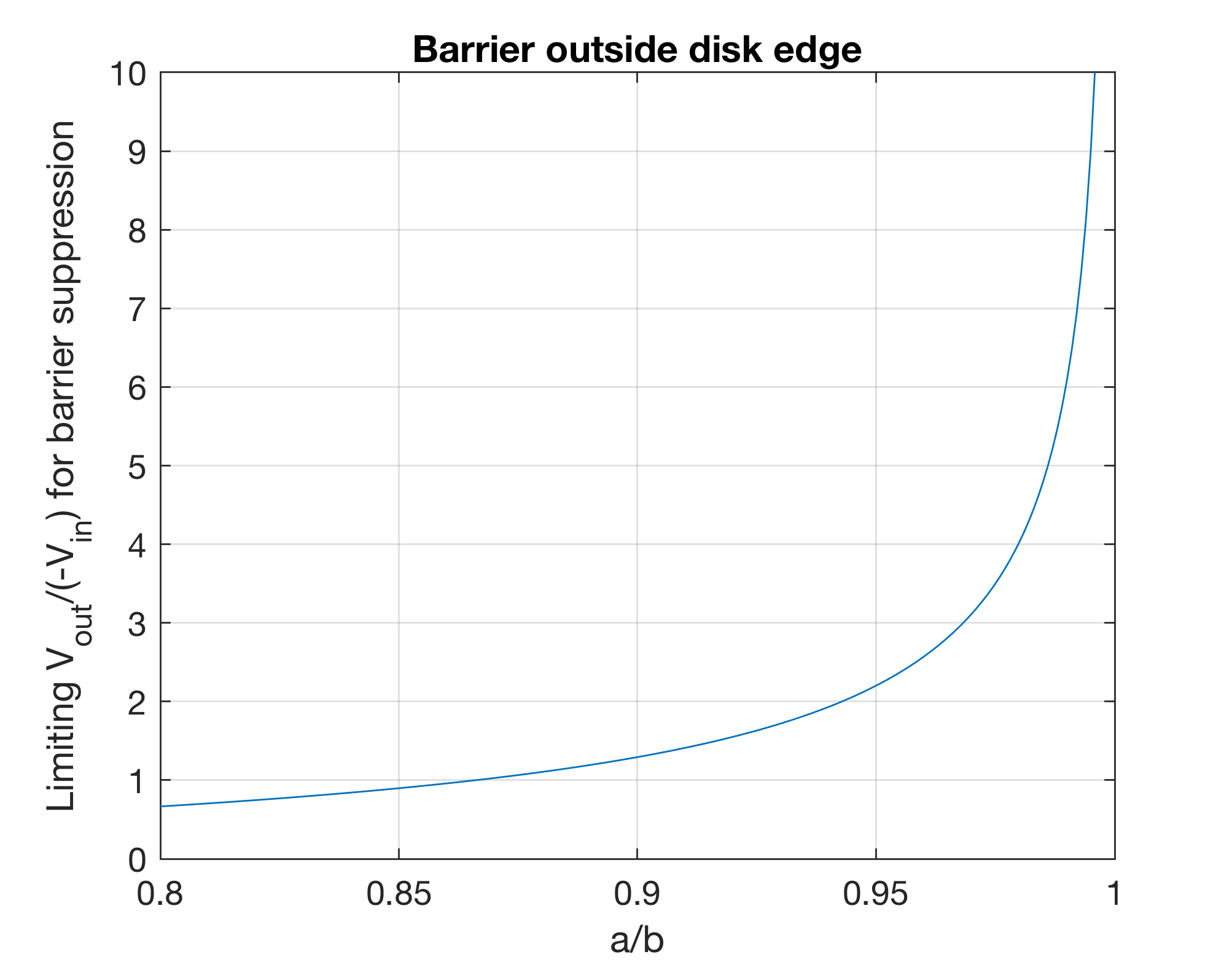}
    \caption{Minimum value of $V\ind{out}/(-V\ind{in})$ for full barrier suppression, as function of the ratio of the radii $a/b$, for a circular disk solar panel in vacuum, where $V\ind{out}$ is positive.}
    \label{fig:barrier_limit}
\end{figure}

In this situation, the general barrier limit Eq.~(\ref{eq:barrier_limit}) takes the form
\begin{align}\label{eq:barrier_limit_rho}
    \frac{V\ind{out}}{-V\ind{in}} =  \frac{1}{\sqrt{1-\left(\frac{a}{b} \right)^2}} - 1.
\end{align}
We plot this limit condition in Figure~\ref{fig:barrier_limit}. For a realistic representation of the \emph{Rosetta} solar panels, $a/b$ should be in the upper end of the plotted range. While the values grow large as the outer ring becomes narrow ($a/b$ close to 1), they are still much more modest than the corresponding ratio in Figure~\ref{fig:z_barrier_limit}, a combined effect of the ring having much larger area then the central circle of similar width and of the circle being exposed to space at the edge of the solar panels with no grounded elements surrounding it. For the value $a/b = 0.98,$ we get a limiting voltage ratio around $-4$. This means that the approximate maximum bias voltage $V\ind{B} = +75$~V (note that $V\ind{B} = V\ind{out}-V\ind{in}$) would be sufficient to attract cold electrons if the spacecraft potential ($V\ind{in}$) is not more negative than $-15$~V. However, this is only a vacuum model and we may expect the shielding provided by the plasma would lower the barrier height and so increase the efficiency of the solar panel edge as a driver of the spacecraft potential.  In Section~\ref{sec:spis_disks}, we show that this actually is the case.

Instead of applying the general condition of Eq.~(\ref{eq:barrier_limit}), we could consider that a barrier in the disk plane means that the potential Eq.~(\ref{eq:Vrho}) has a local maximum. By setting $\mathrm{d}\Phi(\rho,0)/\mathrm{d}\rho=0$, we then obtain the barrier position $\rho_1$ from
\begin{align} \label{eq:barrier_pos}
    \rho_1^2 = \frac{V\ind{in} b a^2}{V\ind{in} b + [V\ind{out}-V\ind{in}] \sqrt{b^2-a^2}}.
\end{align}
Requiring $\rho_1^2>0$ results in the condition Eq.~(\ref{eq:barrier_limit_rho}), but we also find the expression Eq.~(\ref{eq:barrier_pos}) useful as such in Section~\ref{sec:spis_disks}.

Figure~\ref{fig:sherman_contours} shows equipotentials and field lines for this configuration for similar bias values as in Figure~\ref{fig:sherman_contourf_button} (but to much larger distance, that is, four times the disk radius $b$). The barrier can be seen for the two lowest bias cases, where all field lines to infinity connect to the main solar panel area. In the two high-bias cases the barrier is gone and, as discussed in Section~\ref{sec:sherman}, this means that only field lines from the solar panel edge connect to infinity, suggesting that there is an efficient collection of plasma electrons on this edge. Another consequence is that all field lines to the main solar panel area now originate at the biased edge, suggesting a strong suppression of solar panel photoemission. To find if this indeed is the case, we must turn back to the simulations.

\section{Simulation of concentric disks} \label{sec:spis_disks}

\begin{figure}
        \includegraphics[width=1.0\linewidth]{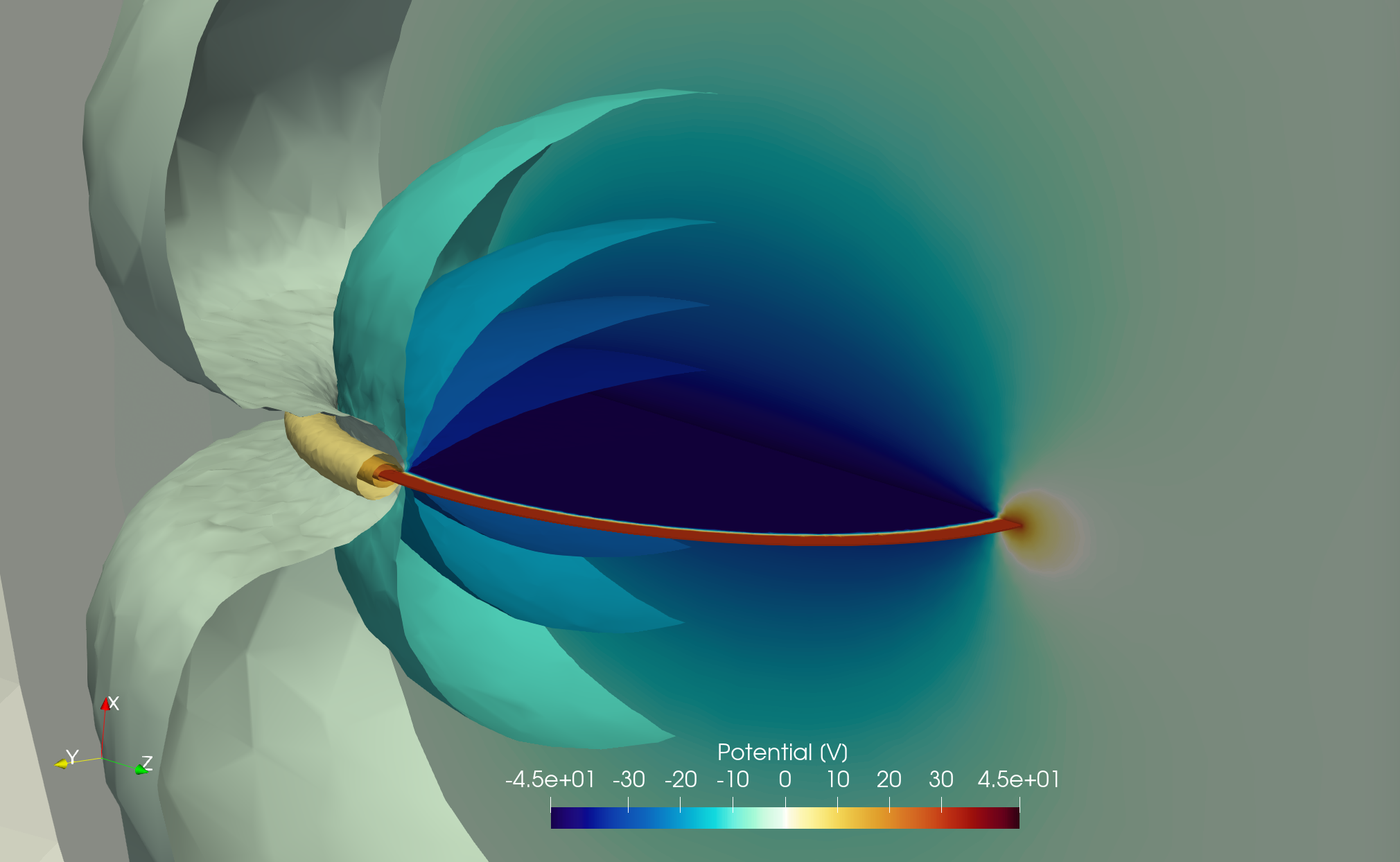}
    \caption{3-D Visualisation of electrostatic potential structure for the SPIS concentric disks ($a=1.17$~m , $b=1.25$~m) simulation with $V_B = +75$~V, coloured by electrostatic potential. To illustrate the potential in the volume, we plot the potential along the X-Z plane, as well as 10 equipotential surfaces from -30~V to +25~V, cut in the X-Y plane. The potential of the outer disk is -45.1~V and 29.9~V for the inner disk.}
    \label{fig:spisdisk}
\end{figure}

To test and extend the validity of the analytical model in Section~\ref{sec:sherman} to incorporate a plasma (and Debye shielding), we simulate two concentric disks at different bias potentials from the ground (0 and +75~V for the inner and outer disk, respectively) in SPIS. We take the models specified in Figure~\ref{fig:disc-sketch} with a disc thickness of 5~mm and use the same environment and materials as specified in Table~\ref{tab:table} with a PIC treatment of all particles and simulate until the spacecraft potential converges. To improve statistics for the electrostatic potential in the volume, we utilise the symmetry of the problem in three dimensions (seen in Figure~\ref{fig:spisdisk}). For plotting the potential along the $z$ axis, we divided the axis into intervals of length d$z$. For each $z$ value, we calculated the median value of the potential for all points between the planes defined by $z$ and $z$+d$z$  (as well as -$z$ and -($z$+d$z$)) which lie within $\rho/a =0.2$. For the plot of potential vs $\rho$, we did the same for a ring-like volume containing all points between a cylinder of thickness d$\rho$ around $\rho$ and within 2 degrees of the $z = 0$ plane.

\begin{figure}
        \includegraphics[width=1.0\linewidth]{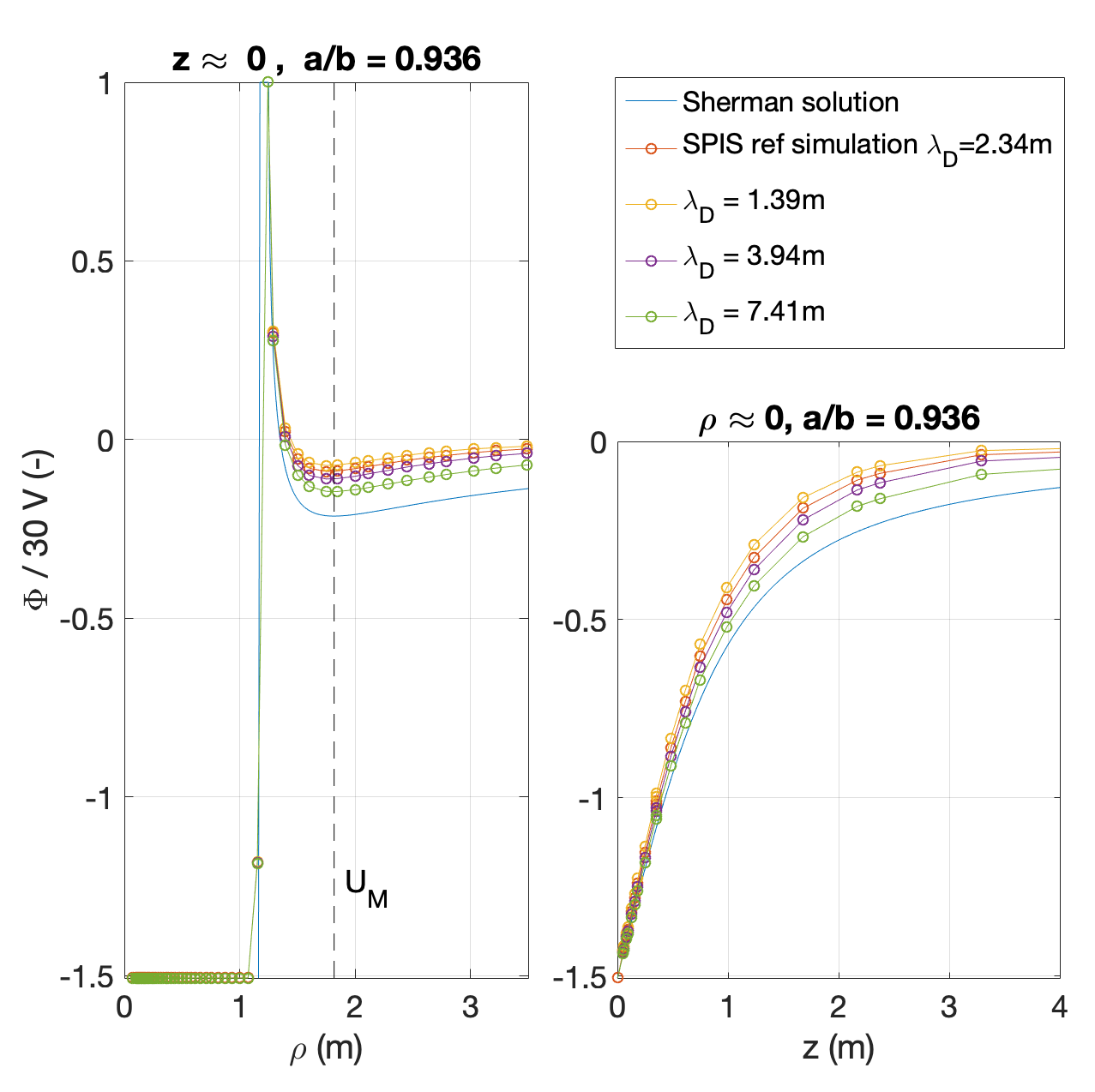}
    \caption{Electrostatic potential normalised by potential of the positive outer ring (+30~V) along cylindrical axes, $\rho$  (\textbf{left}) and z (\textbf{right}). For the vacuum case (blue line), the ring radii fraction, \textbf{a,} is identical to the SPIS model, and the absolute potential of the disks are taken from the output of the reference SPIS simulation (red circles). The associated analytical solution for the barrier potential, $U\ind{M}$ , is marked with a dashed black line. The other SPIS simulations were simulated at identical conditions except with fixed potentials on the rings, and with different plasma Debye lengths (modulated by changing the plasma density or electron temperature in each simulation).}
    \label{fig:spisvssherman}
\end{figure}

\begin{figure}
        \includegraphics[width=1.0\linewidth]{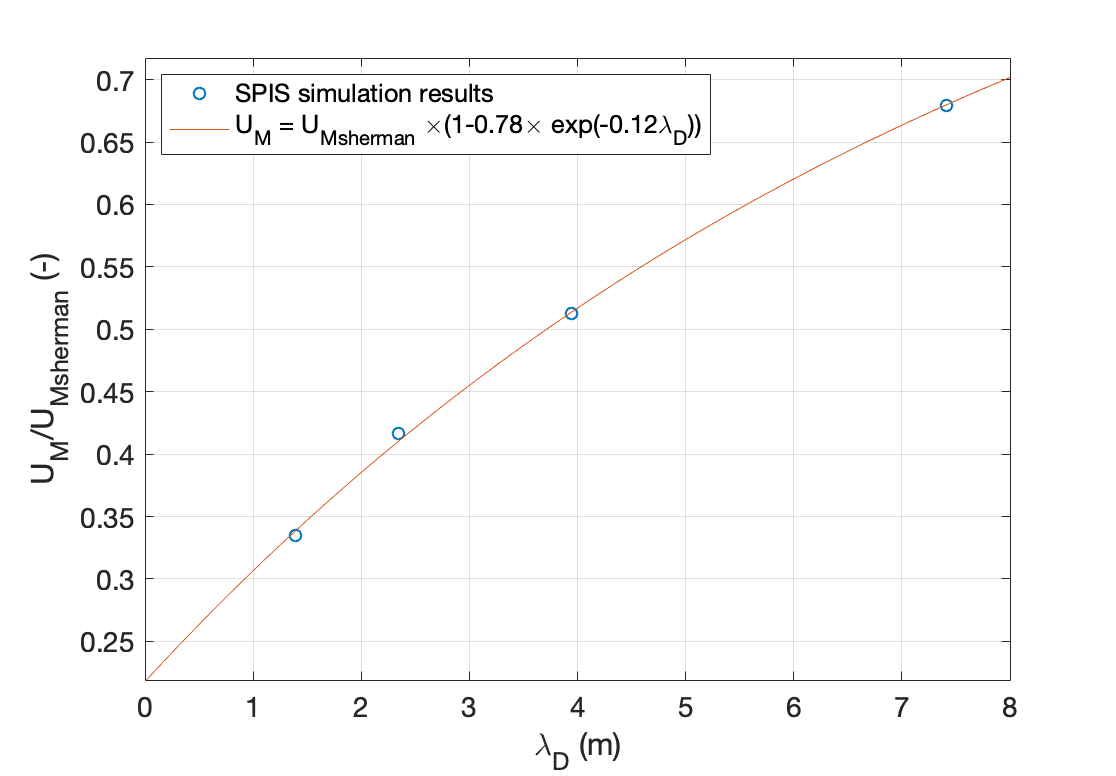}
    \caption{Barrier potential for four SPIS PIC simulations at different Debye lengths with the same disk potentials, divided by the analytic vacuum model result \citep{Sherman1971a}. The fitted model is plotted in red.}
    \label{fig:um_model}
\end{figure}

Comparing the SPIS reference simulation to the vacuum case in Figure~\ref{fig:spisvssherman}, we find a potential barrier at the exact same position (as far as our SPIS simulations resolution allows) as our analytical model predicts. The absolute potential of the barrier $U\ind{M}$ is, unsurprisingly, smaller, as the plasma  would screen all non-zero potentials via Debye shielding.
The SPIS also allows us to run a series of simulations where we can change the Debye length of the plasma in the volume without changing the floating potential of the spacecraft (as changing the density or temperature of the plasma would undoubtedly shift the balance of currents to the spacecraft). These results are also shown in Figure~\ref{fig:spisvssherman}. The potential in the volume is more efficiently damped when moving away from the spacecraft as $\lambda\ind{D}$ decreases and we can plot the fractional departure from the Sherman analytical model at the position of the barrier potential vs $\lambda\ind{D}$ in each simulation in Figure~\ref{fig:um_model}. Using the method of a least-squares fit to an appropriate model, we find that
\begin{equation}\label{eq:um_spis_model}
    U\ind{M} = U\ind{M}^{\dagger}\left(1-0.78\exp{\frac{-\lambda_D}{8.3\text{~m}}}\right),
\end{equation}
  where $U\ind{M}$ is the SPIS barrier potential, $U\ind{M}^{\dagger}$ is the barrier potential  in the vacuum solution, and $\lambda_D$ is the Debye length. This model has the limit $U\ind{M} \lim\limits_{\lambda_D\rightarrow\infty}= U\ind{M}^{\dagger}$ as expected for a vacuum. This particular model clearly does not describe the limit of very short Debye lengths correctly as $U\ind{M}$ here should go to zero, but it does well in describing the parameter range of our simulations and the transition to vacuum conditions.

\section{Spacecraft potential model} \label{sec:finalRosettamodel}

\subsection{Model of current collection with barrier potential}

At distances from a negatively charged object sufficiently large so that the potential has decayed by a few times, $k\ind{B}T\ind{e}/e$, it follows from Liouville’s theorem that the electron density equals the ambient plasma density reduced by a Boltzmann factor~\citep{laframboise_probe_1973}.

As the potential is repelling from infinity up to the barrier, the plasma density at this point should be
\begin{equation}\label{eq:boltzmannfac}
    n\ind{eM} = n \exp{\left(\frac{eU\ind{M}}{k\ind{B}T\ind{e}}\right)}.
\end{equation}
\citet{Olson2010a} therefore proposed that the electron current, $I\ind{ep}$ , to a positive spherical probe within the negative barrier potential from a larger sphere some distance away is given by the standard orbital motion limited expression \citep{mott-smith_theory_1926} with the source density reduced by the Boltzmann factor at the barrier, viz.
\begin{equation}\label{eq:olsson_part1}
    I\ind{ep} = I\ind{e0} \exp{\left(\frac{eU\ind{M}}{k\ind{B}T\ind{e}}\right)}\left(1+\frac{e\left(U-U\ind{M}\right)}{k\ind{B}T\ind{e}}\right),
\end{equation}
where $I\ind{e0} = Ane\sqrt{\frac{k\ind{B}T\ind{e}}{2\pi m\ind{e}}}$ and it is further assumed that the current attraction is governed not by the absolute potential of the anode with respect to a plasma at infinity, but of the difference between the barrier potential and the anode. An electron passing the barrier may very well complete an orbit around the probe sphere and leave again, so the spherical probe form assumed by \citet{Olson2010a} does not seem unrealistic. However, our case of an attracting ring on the edge of a repelling disk is quite different. An electron passing the barrier potential seen in Figure~\ref{fig:sherman_contours}(ab) and Figure~\ref{fig:spisdisk} cannot encircle the anode along the disk edge in the poloidal direction and is efficiently focused toward the edge by the repelling field lines from the main solar panel. This is verified by our PIC simulations in which the current to the anode does not fit any spherical OML expression but is best described by the current collection to a plate of equal area. As such, Eq.~(\ref{eq:olsson_part1}) is simplified to
\begin{equation}\label{eq:olsson}
    I\ind{e} =I\ind{e0} \exp{\left(\frac{eU\ind{M}}{k\ind{B}T\ind{e}}\right)}.
\end{equation}

\subsection{Spacecraft potential from a current balance}\label{sec:OML_Olsson}

\begin{figure*}
    \includegraphics[width=1.0\linewidth]{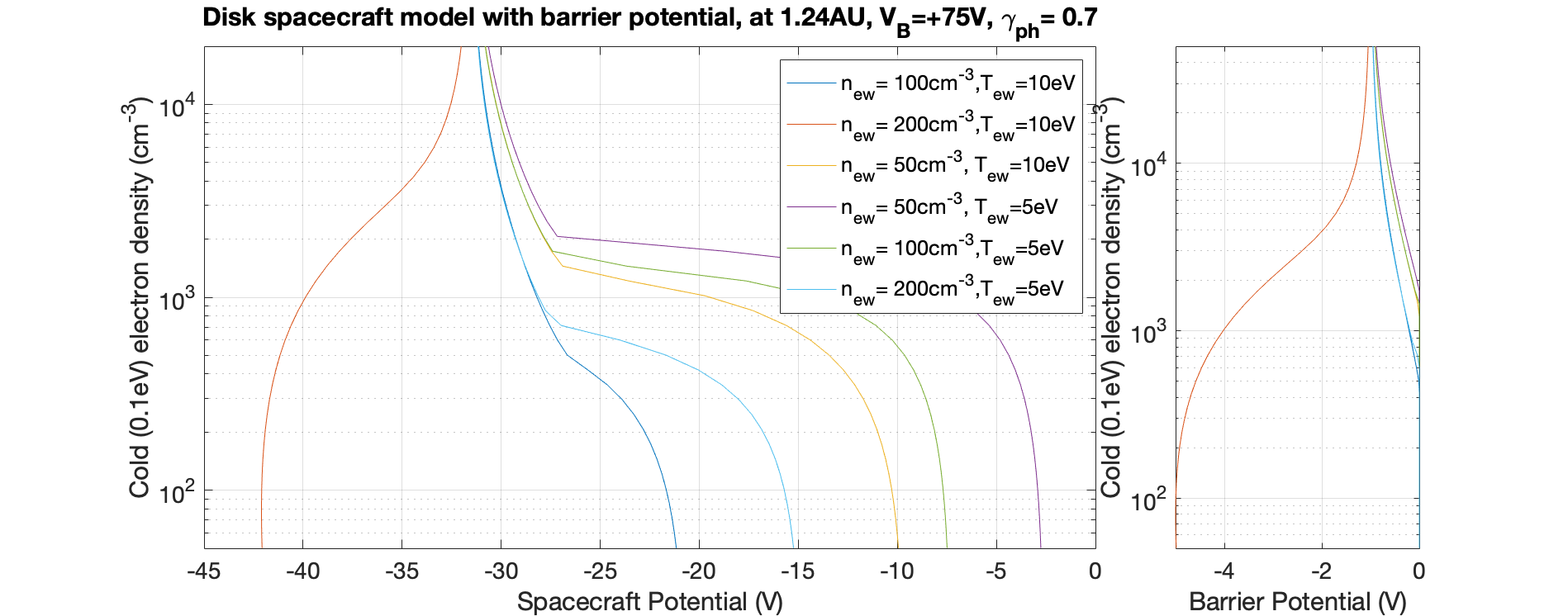}
    \caption{Cold electron density vs Spacecraft potential model solution (\textbf{left}) and barrier potential (\textbf{right}) for various cometary plasma parameters.}
    \label{fig:SOS}
\end{figure*}

\begin{figure*}
        \includegraphics[width=1.0\linewidth]{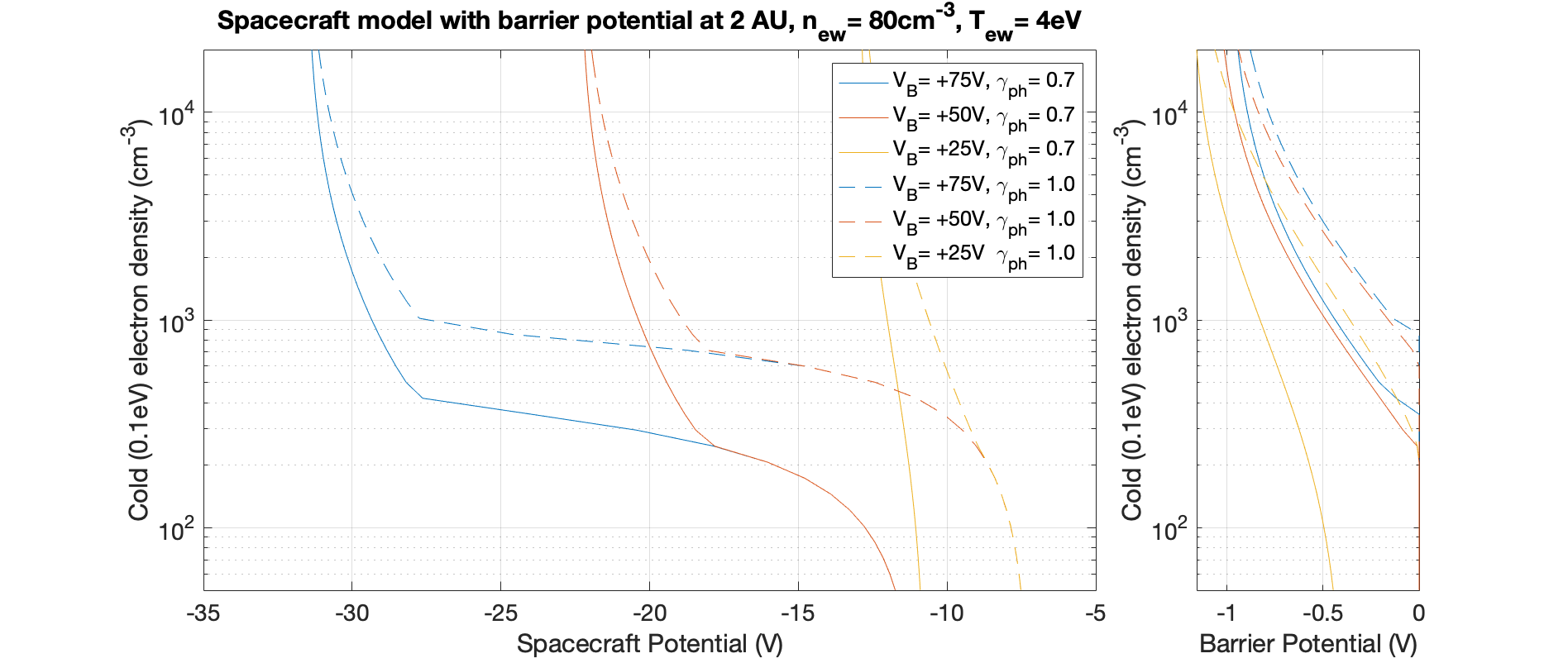}
    \caption{\textbf{Left:} Spacecraft potential model solution for a solar panel vs $n\ind{ec}$ at 2~AU, $n\ind{ew}$ = 80cm$^{-3}$, $T\ind{ew}=4$~eV at various bias potentials. \textbf{Right:} Barrier potential in the model vs $n\ind{ec}$.}
    \label{fig:SOSVb}
\end{figure*}

Bringing it all together, we can estimate the electron current to positive biased elements in a plasma inside a barrier potential (if any) using Equations~(\ref{eq:Vrho}), (\ref{eq:barrier_pos}), (\ref{eq:um_spis_model}), and (\ref{eq:olsson}).
Along with the expression for electron current to a negative surface in OML \citep{laframboise_probe_1973}, we find the general expression for the electron current to a charged solar panel within a barrier potential to be
\begin{equation}
I\ind{e} =
        \begin{cases}\label{eq:all_elec14}
        I\ind{e0} \exp{\left(\frac{ e U\ind{M}}{ k\ind{B}T\ind{e}}\right)} & \text{for } U >  U\ind{M}  \\
    I\ind{e0} \exp{\left(\frac{ e U}{ k\ind{B}T\ind{e}}\right)}  & \text{for } U \le  U\ind{M}. \\
        \end{cases}
\end{equation}
The ion current can be shown \citep{sagalyn_measurement_1963} to be
\begin{equation}\label{eq:ion}
I\ind{i} =
        \begin{cases}
        - I\ind{i0}\left(1 - \frac{ e U}{ E\ind{i}}\right) & \text{for } U < E\ind{i}/e \\
     0 & \text{for } U > E\ind{i}/e, \\
        \end{cases}
\end{equation}
where $E\ind{i}$ is the ion energy, and $I\ind{i0} = Ane\sqrt{2E_i/m_i}$. From the lessons learned in Section~\ref{sec:pillow}, we also reduce the photoemission on the spacecraft by a positive factor, $\gamma\ind{ph} \le 1$, to simulate all interconnectors and bus bars scattered over the solar panels that reduce the net photoelectron production for the spacecraft. In this way, we can simplify the photoemission current expression in \citet{grard_properties_1973} for both surfaces to
\begin{equation}\label{eq:ph}
I\ind{ph} =
        - \pi r\ind{a}^2j\ind{ph0}\left(\frac{1}{d^2\ind{AU}}\right) \gamma\ind{ph},
\end{equation}
where $d\ind{AU}$ is the heliocentric distance in AU, and $r\ind{a}$ is the radius of the inner disk.

We find the equilibrium (spacecraft) potential for our solar panel when all currents to an inner disk at potential $V\ind{S} < U\ind{M}$ and an outer disk at potential $V\ind{S}+V\ind{B}$, sum up to zero. After some rearranging to isolate $V\ind{S}$ from Eq.~\ref{eq:all_elec14}, we find
\begin{equation} \label{eq:final_model}
    V\ind{S}= \frac{k\ind{B}T\ind{e}}{e}\ln\left(\frac{-I\ind{ph} -I\ind{i}^a(V\ind{S})-I_{tot}^b(V\ind{S}+V\ind{B},U\ind{M})}{I\ind{e0}^a}  \right),
\end{equation}
where $I_{tot}^{b}$ is the sum of the currents $I\ind{e}^b$ and $I\ind{i}^b$ as a function of potential (and barrier potential), and we use a superscript to separate terms for the inner and outer disks of radius $a$ and $b$, respectively. %

Now we have all the tools for predicting the current to the \emph{Rosetta} solar panel without the need for computationally costly 3D PIC simulations. In an iterative solution of Eq.~(\ref{eq:final_model}) of all currents to all surfaces on a concentric disk model of ten solar panels and a simple conductive and photo-emitting \emph{Rosetta} SC box of 2x2x2.5m, where we insert also a cold (0.1~eV) electron component that is otherwise computationally costly to simulate, we find the equilibrium spacecraft potential and the barrier potential. We plot the results in Figure~\ref{fig:SOS}.

As we increase the cold electron component, we observe a significant negative charging up until a barrier potential is formed around $V\ind{S} \approx -27$~V. Beyond this potential, or beyond the creation of a barrier potential, the electron density dependence on spacecraft potential tapers of rapidly as cold electrons can no longer reach the spacecraft (even when the net potential of the biased elements is still $\approx +35$~V). When comparing this result to Figure~\ref{fig:vsc_and_gilet}, which prompted this study, we find an explanation for both the highly negative spacecraft potential and the strong dependence on cold electron density versus spacecraft potential below -25~V. As we move to larger heliocentric distances in Figure~\ref{fig:SOSVb}, we shift the curve downwards as the photoemission current decreases everywhere and find a linear trend in the same regions of densities and potentials as in Figure~\ref{fig:vsc_and_gilet}. In reality, the potential of the positive elements around the edge that we base our disk model upon should be distributed on some potential between +0.7 and +78~V and we see in Figure~\ref{fig:SOSVb} that for low bias potentials $V\ind{B}$, the cold electron current has no coupling to the spacecraft or loses coupling even for low spacecraft charging as a potential barrier develops, indicating that moderate (absolute) positive potentials on the interconnects will have little effect on the \emph{Rosetta} spacecraft current system.

In reality, \emph{Rosetta} is not simply a set of solar panels and an ITO-coated spacecraft box but, rather, these represent the principle surfaces for photoemission and current collection and, as such, we believe that the current balance would only be slightly perturbed by incorporating a more realistic \emph{Rosetta} spacecraft body. A more accurate shape model for the solar panels would increase the complexity of the barrier potential shape and the subsequent current collection to all surfaces, but it could be expected to have the same general behaviour in terms of the creation of barrier potentials and current collection to the positively biased surfaces. Introducing a flowing plasma and possible wakes associated with this flow may also alter the details of the balances, but wake effects should mostly be weak: those that are typically observed ion flow directions -- between radially outward from the comet and anti-sunward \citep{bercic_cometary_2018} -- in combination with the \emph{Rosetta} trajectory around 67P -- with a mostly terminator orbit and its solar panel length axis along the direction perpendicular to the nucleus and the panels themselves normal to the sun -- minimises wake effects on the solar panel front surface. We cannot be certain that all our simplifications are valid when moving to a more realistic model, but Figures~\ref{fig:vsc_and_gilet} and \ref{fig:SOSVb} show that we have found a candidate model that describes the general evolution of the spacecraft potential and current collection behaviour of cold electrons to \emph{Rosetta} during its mission. Regardless of geometry, this current collection behaviour seen on \emph{Rosetta} can only be represented by a positively biased conductor with sufficient bias to attract electrons beyond $V\ind{S}<-20~$V and significant surface area to both overcome barrier potentials from surrounding surfaces as well as yield significant cold electron current. We also note that we do not see a decoupling of cold electrons to spacecraft potential in Figure~\ref{fig:vsc_and_gilet} even at the most negative potentials, which indicates that the positively biased surfaces are always (at least for this set of measurements taken between January to September 2016) large or positive enough that no negative barrier potential is formed. Therefore, as shown in Figure~\ref{fig:sherman_contours}, the solar array likely appears as net positive surfaces for a charged particle far away from the spacecraft.

\begin{figure}
        \includegraphics[width=0.9\linewidth]{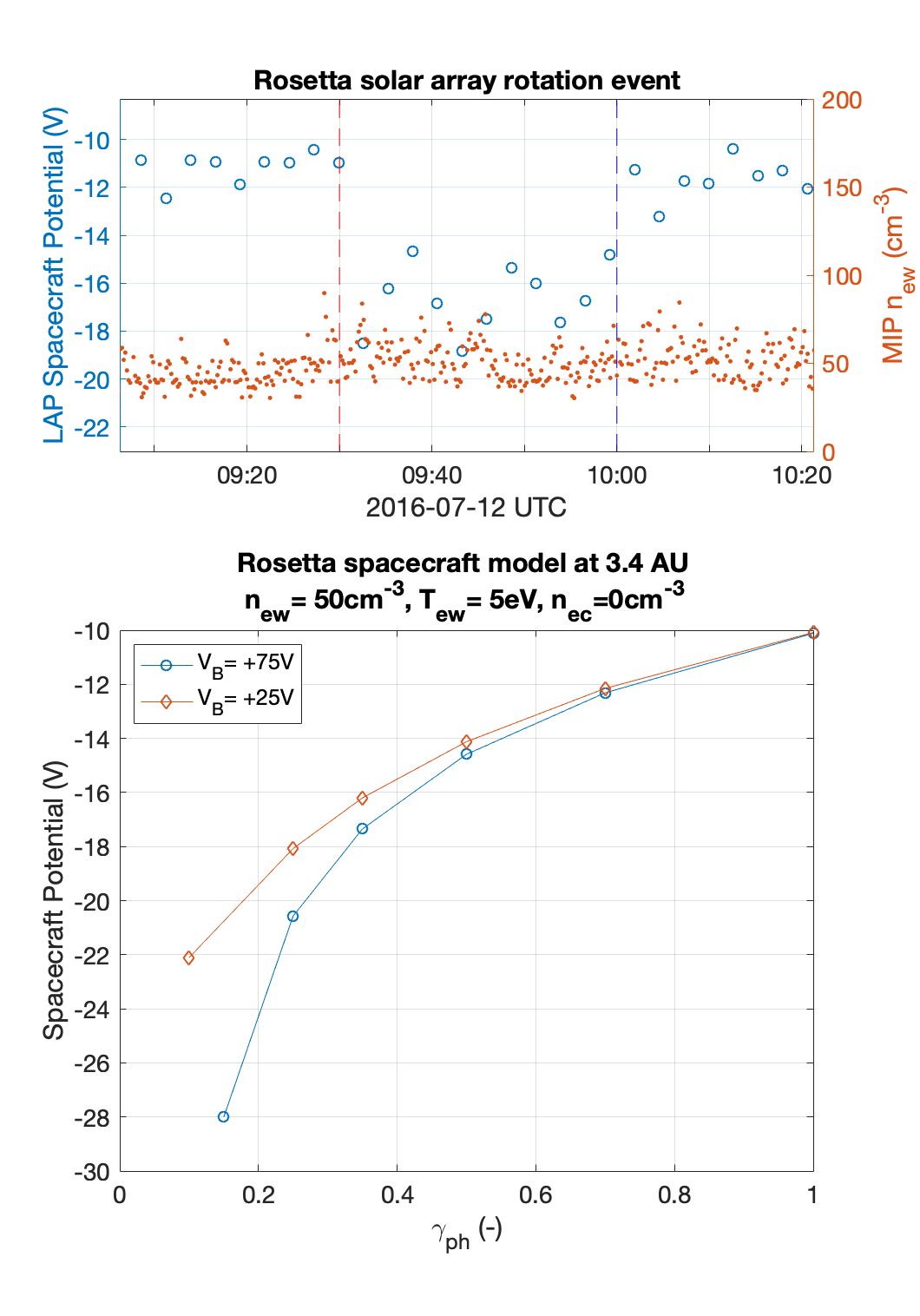}
    \caption{\textbf{Top:} LAP spacecraft potential (blue circles) and MIP electron density (red dots) during a rotation of the Rosetta solar array of 60$\deg$ from the sun, for which two dashed lines indicate the start (red) and end (blue) of the test. \textbf{Bottom:} Spacecraft potential result at various $gamma_{ph}$ and $V_B$ from an adaptation of our solar panel model to a \emph{Rosetta} spacecraft box with ten panels, for which $n\ind{ew}$ = 50$^{-3}$,$T\ind{ew}=5$~eV, at 3.4~AU with no cold electrons present.}
    \label{fig:sa_rotation}
\end{figure}

On 12 July 2016, from 09:30 to 10:00 UTC, \emph{Rosetta} conducted a solar array power test composed of a 60~$\deg$ rotation of the solar array from the sun, thus reducing the available solar power (and photoemission on the solar array) by 50 percent, along with, perhaps, a decrease of $V_B$ on the solar array anodes, as more strings would be at the maximum power point. At this moment, LAP registered a drop in spacecraft potential, from $\approx -11.5$~V to  $\approx -17$~V, plotted in Figure~\ref{fig:sa_rotation}, which is consistent with a spacecraft that is photo-emitting less. As the plasma parameters were otherwise  relatively stable (no detected cold electron population, $T_{ew} \approx 5$~eV), this is a rare opportunity where we can compare these measurements to our model for various parameters in an attempt to constrain $\gamma_{ph}$ and possibly $V\ind{B}$. In Figure~\ref{fig:sa_rotation}, we see that a decrease of 50 percent of photoemission, which in our model represents a 50 percent decrease in $\gamma_{ph}$ (from before the rotation) is compatible with our measurements  for $\gamma_{ph} \gtrapprox 0.4$ in both absolute potential values and the relative potential drop, of which we suspect the latter to be a more relevant parameter, and $\gamma_{ph} = 0.8$ gives us the best fit. What is also apparent in Figure~\ref{fig:sa_rotation} is that we cannot put strong constraints on the positive bias $V_B$, as there are no detectable cold electrons.

\section{Conclusions}

In our investigation of the correlation of the LAP measured \emph{Rosetta} Spacecraft potential to the MIP measured densities and characteristic temperatures of two detected cometary electron populations, we find the spacecraft potential to depend more on electron density (particularly cold electron density) and much less on electron temperature than expected in the high flux of thermal (cometary) ionospheric electrons.

To investigate the current to the positively biased borders on the front-side panels of the \emph{Rosetta} solar array, we first apply an analytical model to obtain the potential surrounding two concentric disks at different potentials in a vacuum. Comparing the result to 3D PIC SPIS simulations and constructing a simple model bridging the two, we arrive at a system of equations that can readily explain the strong relationship between the (highly negative) \emph{Rosetta} spacecraft potential and an observed cold (0.1~eV) electron population that sometimes dominate the \emph{Rosetta} electron environment even when barrier potential effects are considered. We find an explanation for the highly negative charging on the spacecraft, the seemingly poor coupling of electron temperature to spacecraft potential, and the observed log-lin relationship of electron density to spacecraft potential that is used in our analysis of LAP and MIP data to retrieve electron density estimates published on AMDA (\url{http://amda.cdpp.eu/}) and soon on the ESA Planetary Science Archive.

To mitigate spacecraft charging on planetary plasma missions in the future, especially in dense, cold environments where low-energy plasma particles are of particular scientific importance, we suggest an inversion of polarities (i.e. setting spacecraft ground as the anode) in the solar array design, which would drastically reduce the current drawn from the plasma. This approach has been taken on ionospheric spacecraft like Atmospheric Explorer~C and Swarm and been shown to result in a stable and slightly negative potential \citep{Samir1979a,Zuccaro1982a}. Increased efforts to insulate positively biased conductors, where particular attention should be directed to areas close to edges of structures, would also help reduce spacecraft charging and enable more sensitive plasma measurements of the coldest plasma populations.

\begin{acknowledgements}
      \emph{Rosetta} is an ESA mission with contributions from its member states and NASA. This work would not have been possible without the collective efforts over a quarter of a century of all involved in the project and the RPC. We are also grateful to everybody who has worked on the SPIS software, at ONERA, ARTENUM and elsewhere, and to ESA for supporting this highly valuable package. This research was funded by the Swedish National Space Agency under grant Dnr 168/15. The cross-calibration of LAP and MIP data was supported by ESA as part of the Rosetta Extended Archive activities, under contract 4000118957/16/ES/JD. %
\end{acknowledgements}

\bibliographystyle{aa}
\bibliography{References_all.bib}

\end{document}